\documentclass[twocolumn,longabstract,numberedappendix,twocolappendix]{aastex63}
\usepackage{amsmath,amssymb,commath}
\usepackage[displaymath,mathlines]{lineno}
\usepackage{tikz}
\usetikzlibrary{matrix,positioning,decorations.pathreplacing}
\usepackage{afterpage}
\usepackage{rotating}
\usepackage{graphicx}
\usepackage{color}
\usepackage[normalem]{ulem}
\usepackage{subfloat}
\usepackage{verbatim}
\usepackage{makecell}

\usepackage{pifont}
\newcommand{\cmark}{\textcolor{teal}{\ding{51}}}
\newcommand{\xmark}{\ding{55}}%

\received{January 1, 0000}
\revised{January 1, 0000}
\accepted{January 1, 0000}
\submitjournal{ApJ}

\shorttitle{SFFT}
\shortauthors{Hu et al.}

\graphicspath{{./}{}}

\begin{document}

\title{Image Subtraction in Fourier Space}

\correspondingauthor{Lifan Wang}
\email{lifan@tamu.edu}

\author[0000-0001-7201-1938]{Lei Hu}
\affiliation{Purple Mountain Observatory \\
Nanjing 210023, People's Republic of China}
\affiliation{School of Astronomy and Space Sciences \\
University of Science and Technology of China \\
230026 Hefei People's Republic of China}

\author[0000-0001-7092-9374]{Lifan Wang}
\author[0000-0003-3021-4897]{Xingzhuo Chen}
\author[0000-0002-1376-0987]{Jiawen Yang}
\affiliation{George P. and Cynthia Woods Mitchell Institute for Fundamental Physics \& Astronomy, \\
Texas A. \& M. University, Department of Physics and Astronomy, 4242 TAMU, College Station, TX 77843, USA}

\begin{abstract}

Image subtraction is essential for transient detection in time-domain astronomy. The point spread function (PSF), photometric scaling, and sky background generally vary with time and across the field-of-view for imaging data taken with ground-based optical telescopes. 
Image subtraction algorithms need to match these variations for the detectionof flux variability. An algorithm that can be fully parallelized is highly desirable for future time-domain surveys. Here we show the Saccadic Fast Fourier Transform (SFFT) algorithm for image differencing. SFFT uses $\delta$-function basis for kernel decomposition, and the image subtraction is performed in Fourier Space. This brings about a remarkable improvement of computational performance of about an order of magnitude compared to other published image subtraction codes. SFFT can accommodate the spatial variations in wide-field imaging data, including PSF, photometric scaling, and sky background. However, the flexibility of the $\delta$-function basis may also make it more prone to overfitting. The algorithm has been tested extensively in real astronomical data taken by a variety of telescopes. Moreover, the SFFT code allows for the spatial variations of the PSF and sky background to be fitted by spline functions.

\end{abstract}



\section{Introduction} \label{sec:intro}

Since \citet{Zwicky1964}, variable source identification from astronomical observations has played an indispensable role in time-domain astrophysics. 
However, the time and spatial variations of point spread function (PSF), the noises in the sources and the sky backgrounds, the optical distortions of the observing facilities, and many other effects barricade rapid and robust transient detection over wide sky areas. 
An efficient algorithm for calculating image differences is crucial in a broad range of astronomical observations of today. Examples include transient searches, as well as microlensing \citep{Mao91} and pixel lensing \citep{Crotts92,Baillon93}. 
The techniques of image differencing and co-addition are also of central importance to the next-generation survey telescope, such as the Legacy Survey of Space and Time (LSST) \citep{LSST19} and the Nancy Grace Roman Space Telescope.

To solve the PSF discrepancies between a pair of images taken at different epochs of the same field, \citet{Phillips_Davis_95} and \citet{Tomaney_Crotts_96} introduced a convolution kernel to match PSFs from one image to the other. 
They use a simple deconvolution solution to determine the kernel by calculating the ratio of the Fourier transform of a bright star or a pre-computed PSF of each image. 
However, this approach cannot guarantee optimal subtractions since the division in Fourier space is prone to numerical instability, especially in the noise-dominated high frequencies domain \citep{AL98,ZOGY}.
Given that the PSFs are only modeled from the isolated stars with sufficient signal-to-noise ratios, the method has difficulty in using all statistically valid pixels in kernel determination, which makes it challenging to handle very crowded fields as in microlensing surveys.
With the same goal of finding the matching kernel, \citet{Kochanski96} moved the first step towards kernel determination without any direct knowledge of PSFs. They solved the problem by a least-squares fitting on the image-pair under consideration instead of extracting the light profiles. Nevertheless, the required computing time of the non-linear fitting introduced by \citet{Kochanski96} is generally formidable \citep{AL98}.

Soon after \citet{Kochanski96}, a ``fork in this road" was signposted by the pioneering paper of \citet{AL98}.
They decomposed the convolution kernel into a basis of functions, converting the problem to be a linear least-squares question. 
To accommodate the varying PSFs across the image, \citet{AL98} modeled a spatially variant convolution kernel that evolves with image coordinates in a polynomial form. Subsequently, \citet{AL00} simplified the calculations for constructing the least-squares matrix, making it possible to fit the kernel spatial variation with a reasonable computing cost. 
In the last two decades, the framework outlined in \citet{AL98} and \citet{AL00} serves as the mathematical foundation upon which several successful transient survey programs have been built \citep[e.g.,][]{iPTF_Cao16,DES18,PANSTARRS19}. As the approach does not require the existence of isolated stars, it has been applied to very crowded fields, such as microlensing observations toward Galactic bulge \citep[e.g.,][]{Sumi_2003,Sumi_2006,Sumi_2013}.

\citet{AL98} provided the essential building blocks of image subtraction, subsequent studies continue to move the field forward, with a focus on improving the kernel basis functions and developing kernel regularization techniques.
The Gaussian basis functions adopted in \citet{AL98} only allow for incomplete expansion of the convolution kernels. As a result, the method is constrained by the bases' intrinsic symmetry and hinders the ability to construct a ``shape-free" kernel \citep{Becker12}. Moreover, one has to configure a variety of hyper-parameters for the kernel basis, e.g., the number and width of the Gaussians. 
In order to achieve sheer kernel flexibility and minimize user-adjustable parameters, \citet{Bramich08} and \citet{Miller08} introduced the delta basis functions (DBFs) for the construction of complete kernel space.
With the DBFs, image subtraction is capable of compensating for the sub-pixel astrometric misalignment through an unconstrained flux redistribution within the scale of kernel size \citep{Bramich08,Becker12}.
\citet{Albrow09} confirmed the compelling advantage of DBFs by showing a distinct improvement in photometry accuracy over the traditional Gaussian basis functions.
\citet{Bramich13} further developed the DBFs-based approach by taking the spatially varying photometric scale into account, aimed to accommodate the 
variation of transparency and airmass across the field in wide-field surveys.

The flip side of the ultimate flexibility of DBFs is that the resulting kernel solution is more susceptible to noise. 
To alleviate the overfitting problem, \citet{Becker12} conceived a Tikhonov regularization by adding a penalty term so that the solution would favor compact and smooth kernels. 
\citet{iPTF_Masci17} adopted an easily implemented way to regularize the kernel with DBFs for iPTF transient discovery. They used 
to solve the least-squares but dropped the eigenvalues with low statistical significance. 
Regularizing the kernel itself is not the unique option to fight against the overfitting problem. \citet{Bramich16} found that using a parsimonious choice of unregularized DBFs might be an even better alternative.

Apart from the methods that stem from \citet{AL98}, a numerically stable approach characterized by cross-convolution was proposed by \citet{Gal08} and \citet{Yuan08}. 
This strategy became more prevalent since \citet{ZOGY} (hereafter ZOGY) presented a closed-form algorithm that yields the optimal subtraction for transient source detection. 
By its design, ZOGY can result in difference images with uncorrelated background noise. With the noise propagation during the image subtraction process, ZOGY claims optimal detection rates of transients with minimal subtraction artifacts.

In addition to the subtraction algorithms mentioned above, machine learning approaches are making their way into the field.
\citet{Sedaghat18} recently suggested the use of deep convolutional neural networks instead of PSF-matching for transient detection. 
\citet{PyTorchDIA21} (hereafter PyTorchDIA) proposed a novel machine learning approach to optimize the convolution kernel by making use of automatic differentiation in PyTorch instead of constructing a least-squares matrix. 

We show in this paper that the entire imaging differencing process can be performed with Fast Fourier Transform (FFT). The FFT can be easily adapted to different computer environments for parallel processing.
Our method, Saccadic Fast Fourier Transform (SFFT), provides a massively parallelizable framework for image subtraction and brings substantial processing speed acceleration when implemented on Graphical Processing Units (GPU).
Apart from the gain in computing cost, SFFT can retain the crucial features that came from the advances in image subtraction since \citet{AL98}, including (1) SFFT employs flexible DBFs as kernel basis functions; (2) SFFT can adequately handle the spatial variations of convolution kernel, photometric scale, and differential background; (3) SFFT does not rely on the availability and distribution of the observed stellar objects.

The paper is structured as the following: Section~\ref{sec:sfft_methodology} provides the mathematical frameworks of SFFT. Section~\ref{sec:sfft_implement} introduces the implementation of SFFT and the details of the software pipeline. Section~\ref{sec:sfft_examples} shows the application of SFFT pipeline to data taken from a variety of telescopes.
Section~\ref{sec:sfft_comput_perf} presents the computing performance of SFFT, with comparisons to other existing image subtraction implementations.
Section~\ref{sec:sfft_limfuture} discusses the limitations of SFFT and plans for future works. All the code about SFFT subtraction is available on Github\footnote{\url{https://github.com/thomasvrussell/sfft}}.

\section{Methodology of SFFT} \label{sec:sfft_methodology}

\subsection{Overview} \label{subsec:m-overview}

As the core engine of transient detection, image subtraction is often the most computation-intensive individual task in the data processing pipeline. 
The ongoing trend of very massive data flow from time-domain surveys is making real-time data reduction increasingly challenging. It motivated us to develop a new tool to relieve the computational bottleneck while reconciling the subtraction performance and the computing cost.

The image subtraction algorithms emanated from \citet{AL98} have been broadly applied in astronomical transient surveys. Our proposed method SFFT is a new member of this category.
The main purpose of the approach aims to perform PSF-matching via image convolution with pixelized kernels formulated as linear combinations of a pre-defined set of basis functions.
The fact that astronomical images from wide-field surveys generally possess non-constant PSF across the entire field of view leads to spatial variations of the convolution kernels. 
Moreover, the spatial varying photometric scale also needs to be taken into account in order to match the images from wide-field surveys.

There have been some attempts to accelerate the calculations involved in image subtraction, either aimed at faster kernel determination or speed-up of the subtraction afterward. 
For image subtraction, constructing the least-squares matrix to solve the convolution kernels is computationally expensive, especially for spatial variant kernels. \citet{AL00} proposed to fit the space-varying kernel on a set of sub-areas of the field, with a simple assumption that the kernel spatial variation within each sub-area can be negligible. The useful strategy has been adopted in the software {\tt\string HOTPANTS}, a widely-used implementation of \citet{AL98}.
Apart from the efficient simplification in computing, the flexibility of using sub-areas has some additional benefits in practice. One may pre-select an optimal set of sub-areas to exclude the sources in observational data that cannot be perfectly modeled by the image subtraction algorithm. Doing so can also avoid the fitting being strongly misled by specific regions of the field (e.g., the brightest and densest regions).

Very recently, \citet{PyTorchDIA21} provided an innovative way (i.e., PyTorchDIA) to bypass the construction of the analytical least-squares matrix. PyTorchDIA finds the kernel solution by optimizing a loss function using the automatic differentiation in PyTorch, which brings a considerable gain in computing cost compared with \citet{Bramich08}.
However, PyTorchDIA is incompatible with kernel spatial variation and can only solve a constant kernel for image subtraction. 
In addition to the efforts on kernel determination, \citet{Hartung12} successfully implemented fast image convolution with spatial-varying kernel on GPUs, which helps accelerate the subtraction after the convolution kernel has been computed.

The strategy of SFFT is different from the methods mentioned above: we present the least-squares question in Fourier space. 
By Parseval's theorem, a least-squares minimization in real space is equivalent to a least-squares minimization in Fourier's domain. 
Given that $\delta$-functions have higher flexibility and can maintain simple forms after Fourier transform, here we adopt the DBFs as the kernel basis functions following \citet{Bramich08} and \citet{Miller08}.
Finally, SFFT allows for spatial variations across the field, whether from PSF, photometric scaling, or sky background.
The following section will show how the calculations for image subtraction can be reduced to be FFTs and element-wise matrix operations.

\subsection{Derivation of Subtraction} \label{subsec:m-subtract}

Given a reference image $R$ and a science image $S$, PSF-homogenization is carried out by convolving $R$ or $S$ with a spatially-varying kernel $K_{x, y}$. Considering that the sky background between the two epochs may change, we introduce an additional term $B$ to model the sky background difference between the two images. The image subtraction problem can be written in real space as the minimization of the difference image $D$, defined as

\begin{equation}
\begin{split}
D(x, y) = S(x, y) & - \int\int dudv R(x-u, y-v)K_{x,y}(u, v) 
\\
& - B(x, y)
\\
= S(x, y) & - (R\circledast K)(x,y) - B(x, y),
\end{split}
\label{eqn:sfft_eq1}
\end{equation}
where the input images $R$ and $S$ are image with dimensions $(N_0, N_1)$, and $x$ and $y$ are indices in the ranges of $[0, N_0-1]$ and $[0, N_1-1]$, respectively. Note that the kernel $K_{x, y}$ is spatially varying, so the integral in the equation above, strictly speaking, is not a convolution, such an integral is denoted by $R\circledast K$. The differential background map $B$ is modeled as a polynomial form following \citet{AL98}, that is,
\begin{equation}
B(x, y) = \sum_{pq} b_{pq} x^{p} y^{q}
\label{eqn:sfft_eq2}
\end{equation}
where the polynomial power indices $p$ and $q$ are in the range of $[0, D_B]$ and $[0, D_B-p]$, respectively.
We follow \citet{Miller08,Hartung12} to decompose the kernel $K_{x, y}$ into the ``shape-free" $\delta$-function basis $\mathcal{K}$. The kernel dimension is assumed to be $(L_0, L_1)$, with $L_0$ and $L_1$ being odd integers, such that $L_0 = 2w_0+1$ and $L_1 = 2w_1+1$. We assume

\begin{equation}
\mathcal{K}_{00}(u, v) = \mathcal{D}_{00}(u, v) = \vec{\delta}(u, v),
\label{eqn:sfft_eq3}
\end{equation}
and,
\begin{equation}
\label{eqn:sfft_eq4}
\begin{split}
\mathcal{K}_{\alpha\beta}(u, v) &= \mathcal{D}_{\alpha\beta}(u, v) - \mathcal{D}_{00}(u, v) 
\\
&= \vec{\delta}(u-\alpha, v-\beta) - \vec{\delta}(u, v)
\end{split}
\end{equation}
where $\mathcal{D}_{\alpha\beta}(u, v) = \vec{\delta}(u-\alpha, v-\beta)$ is the standard Cartesian orthonormal basis, which consists of $\delta$ functions at each kernel pixel, $\alpha$ and $\beta$ are indices in the range of $[-w_0, w_0+1)$ and $[-w_1, w_1+1)$, respectively, u and v are kernel coordinate indices in the range of $[-w_0, w_0+1)$ and $[-w_1, w_1+1)$, respectively, and 
$\vec{\delta}$ is a binary function on integers: $\vec{\delta} (\rho, \epsilon) = 1$ if $\rho = \epsilon = 0$, otherwise $\vec{\delta} (\rho, \epsilon) = 0$ with $\rho$ and $\epsilon$ being any integers.
With such a construction, the sum of the convolution kernel is uniquely determined by the coefficient of the basis vector $\mathcal{K}_{00}$, which simplifies the way we control the photometric scaling through convolution.

The spatial variation of the kernel can be fitted by the sum of the kernel function given above, modulated by a two-dimensional polynomial function to account for their varying contributions across the image field

\begin{equation}
K_{x,y} = \sum_{\alpha\beta} \mathring{A}_{xy\alpha\beta} \mathcal{K}_{\alpha\beta}
\label{eqn:sfft_eq5}
\end{equation}
\begin{equation}
\mathring{A}_{xy\alpha\beta} = \sum_{ij} \mathring{a}_{ij\alpha\beta} x^{i} y^{j},
\label{eqn:sfft_eq6}
\end{equation}
where 
the polynomial power indices $i$ and $j$ are in the range of $[0, D_K]$ and $[0, D_K-i]$, respectively. The spatial variation information of the kernel is encoded in the coefficients $\mathring{a}_{ij\alpha\beta}$, which will be eventually solved by the subsequent linear system during the minimization of the residuals. More specifically, once $\mathring{a}_{ij\alpha\beta}$ is known, we can calculate $\mathring{A}_{xy\alpha\beta}$ for any given pixel coordinate $(x, y)$ following Equation~(\ref{eqn:sfft_eq6}), and further derive the certain kernel associated with the pixel via the expansion on $\delta$-function basis by Equation~(\ref{eqn:sfft_eq5}).

Equations~(\ref{eqn:sfft_eq2}), (\ref{eqn:sfft_eq5}) and (\ref{eqn:sfft_eq6}) can be replaced by more complex functions. The code we have developed includes an option for using B-splines to model the PSF and background level variations. For the B-splines cases, the space-varying kernel $K_{x,y}$ and differential background B(x, y) are modeled as

\begin{equation}
K_{x,y} = \sum_{\alpha\beta} \mathring{A}_{xy\alpha\beta} \mathcal{K}_{\alpha\beta}
\label{eqn:sfft_eq7}
\end{equation}

\begin{equation}
\mathring{A}_{xy\alpha\beta} = \sum_{ij} \mathring{a}_{ij\alpha\beta} B_{i;k,t}(x) B_{j;k,t}(y)
\label{eqn:sfft_eq8}
\end{equation}

and
\begin{equation}
B(x, y) = \sum_{pq} b_{pq} B_{p;k^\prime,t^\prime}(x) B_{q;k^\prime,t^\prime}(y)
\label{eqn:sfft_eq9}
\end{equation}
where $B_{p;k,t}$ ($B_{p;k^\prime,t^\prime}$) are B-spline basis functions of degree k ($k^\prime$) and knots t ($t^\prime$). For simplicity, in this paper we focus only on the performance with polynomial models.

The pixel-to-pixel flux variations of the two images are accounted for by the coefficients $\mathring{A}_{xy00} = \sum_{ij} \mathring{a}_{ij00} x^{i} y^{j}$. If we consider the flux level of the image-pair to be well-calibrated, the constant flux scaling between images requires a constant kernel integral, that is, $\mathring{A}_{xy00} = \mathring{a}_{0000}$. Note that a constant flux scaling was first presented in \citet{AL98}.
Having a constant photometric ratio across the entire field is optional in our program. Like \citet{Bramich13}, SFFT allows for space-varying flux scaling with a polynomial form to accommodate the effect of imperfect flat-field correction or cirrus cloud attenuation. Note that the current SFFT does not disentangle the polynomial degrees of spatial variations accounting for the convolution kernel and the photometric sensitivity.

With abbreviation $T^{\rho\epsilon} (x, y) = x ^{\rho} y^{\epsilon}$ and $R^{\rho\epsilon} = T^{\rho\epsilon}R$ ,where $\rho$ and $\epsilon$ are any integers, Equation~(\ref{eqn:sfft_eq1}) is rewritten as,

\begin{equation}
\label{eqn:sfft_eq10}
\begin{split}
D(x, y)  =  S(x, y) &- \sum_{ij\alpha\beta} \mathring{a}_{ij\alpha\beta} [T^{ij} (R \circledast \mathcal{K}_{\alpha\beta})] (x, y) 
\\
&- \sum_{pq} b_{pq}T^{pq} (x, y).
\end{split}
\end{equation}
The convolution kernel is typically very small in size, at such scale its spatial variation is expected to be negligible locally. 
An approximation (see Appendix~\ref{App:approx}) can be made by moving the polynomial functions describing the spatial variation outside the integral in Equation~(\ref{eqn:sfft_eq1}). This leads to 

\begin{equation}
\label{eqn:sfft_eq11}
\begin{split}
D(x, y)  =  S(x, y) &- \sum_{ij\alpha\beta} \mathring{a}_{ij\alpha\beta} (R^{ij} \circledast \mathcal{K}_{\alpha\beta}) (x, y) 
\\
&- \sum_{pq} b_{pq}T^{pq} (x, y),
\end{split}
\end{equation}

and, 

\begin{equation}
D(x, y) = S(x, y) - \sum_{ij\alpha\beta} \mathring{a}_{ij\alpha\beta} R^{ij} \circ \mathcal{K}_{\alpha\beta} - \sum_{pq} b_{pq}T^{pq}.
\label{eqn:sfft_eq12}
\end{equation}

Note the notation $\circ$ in Equation~$(\ref{eqn:sfft_eq12})$ indicates circular convolution and $\mathcal{K}_{\alpha\beta}$ can be estimated to generate the convoluted source image that best matches the PSF of the reference image across the entire field. 

The $\delta$-function basis we have adopted for $\mathcal{K}$ as shown in equations~(\ref{eqn:sfft_eq3}) and (\ref{eqn:sfft_eq4}) allows for a simple Fourier space representation of the image difference procedure. In Fourier space Equation~(\ref{eqn:sfft_eq12}) becomes

\begin{equation}
\widehat{D} = \widehat{S} - \sum_{ij\alpha\beta} a_{ij\alpha\beta} \widehat{R^{ij}} \widehat{\mathcal{K}_{\alpha\beta}} - \sum_{pq} b_{pq}\widehat{T^{pq}},
\label{eqn:sfft_eq13}
\end{equation}
where the symbols with a hat denote the Fourier transform of the images, $a_{ij\alpha\beta} = N_0N_1 \mathring{a}_{ij\alpha\beta}$ with $N_0$ and $N_1$ being the dimensional of the images in $x$ and $y$, respectively. The Discrete Fourier Transform (DFT) of the basis function has the simple form

\begin{equation}
\begin{tabular}{rll}
$\widehat{\mathcal{K}_{\alpha\beta}}(l, m) = $& $\frac{1}{N_0N_1}({W_{N_0}^{l\alpha}W_{N_1}^{m\beta} - 1})$ & if $\alpha$ or $\beta$ $\neq 0$ \\
  = & $1/{N_0N_1}$ & if $\alpha$ = $\beta$ = 0,
\end{tabular}
\label{eqn:sfft_eq14}
\end{equation}
here $W_{N_0} = e^{-i2\pi/N_0}$ and $W_{N_1} = e^{-i2\pi/N_1}$, with $i$ being the unitary imaginary number.

Now define $G = \widehat{D} \widehat{D}^{*}$, where $^*$ stands for complex conjugate, we find,

\begin{equation}
\label{eqn:sfft_eq15}
\begin{split}
G &= 2 \Re(\widehat{S}) - 2 \sum_{ij\alpha\beta} a_{ij\alpha\beta} \Re\;[\widehat{S}^* \widehat{R^{ij}}\widehat{\mathcal{K}_{\alpha\beta}}] 
\\
&+2 \sum_{ij\alpha\beta} \sum_{pq} a_{ij\alpha\beta}b_{pq} \Re\;[\widehat{R^{ij}} \widehat{\mathcal{K}_{\alpha\beta}} (\widehat{T^{pq}})^*]
\\
&+ \sum_{ij\alpha\beta} \sum_{i^\prime j^\prime \alpha^\prime \beta^\prime} a_{ij\alpha\beta} \widehat{R^{ij}} \widehat{\mathcal{K}_{\alpha\beta}} a_{i^\prime j^\prime \alpha^\prime \beta^\prime} (\widehat{R^{i^\prime j^\prime}})^* (\widehat{\mathcal{K}_{\alpha^\prime \beta^\prime}})^* 
\\
&-2 \sum_{pq} b_{pq} \Re\;[\widehat{S} ^* \widehat{T^{pq}}] + \sum_{pq}\sum_{p^\prime q^\prime} b_{pq} \widehat{T^{pq}} b_{p^\prime q^\prime} (\widehat{T^{p^\prime q^\prime}})^*.
\end{split}
\end{equation}

Our goal of optimal subtraction is to minimize the power of the difference image in Fourier space, as given in Equation~(\ref{eqn:sfft_eq15}).
Let $F = \sum_{l, m} G(l, m)$, the minimization in Fourier space for image subtraction has a simple least-square solution given by $\nabla F = 0$ where the gradients are calculated with respect to all the parameters for image matching, 

\begin{equation}
\begin{split}
\partial F / \partial a_{\bar{i}\bar{j} \bar{\alpha} \bar{\beta}} &= \sum_{lm} \{-2 \Re\;[\widehat{S}^* \widehat{R ^{\bar{i} \bar{j}}} \widehat{\mathcal{K}_{\bar{\alpha} \bar{\beta}}}] 
\\
&+ 2 \sum_{ij\alpha\beta} a_{ij\alpha\beta} \Re\;[(\widehat{R^{ij}})^* (\widehat{\mathcal{K}_{\alpha\beta}})^* \widehat{R^{\bar{i} \bar{j}}} \widehat{\mathcal{K}_{\bar{\alpha} \bar{\beta}}}]
\\
&+ 2 \sum_{pq} b_{pq} \Re\;[\widehat{R^{\bar{i} \bar{j}}} \widehat{\mathcal{K}_{\bar{\alpha} \bar{\beta}}} (\widehat{T^{pq}})^*]\}(l, m) 
\\
&= 0
\label{eqn:sfft_eq16}
\end{split}
\end{equation}

\begin{equation}
\label{eqn:sfft_eq17}
\begin{split}
\partial F / \partial b_{\bar{p} \bar{q}} &= \sum_{lm} \{-2 \Re\;[\widehat{S}^* \widehat{T ^{\bar{p} \bar{q}}}]
\\
&+ 2 \sum_{ij\alpha\beta} a_{ij\alpha\beta} \Re\;[(\widehat{R^{ij}})^* (\widehat{\mathcal{K}_{\alpha\beta}})^* \widehat{T^{\bar{p} \bar{q}}}] 
\\
&+ 2 \sum_{pq} b_{pq} \Re\;[(\widehat{T^{pq}})^* \widehat{T^{\bar{p} \bar{q}}}]\} (l, m) 
\\
&= 0
\end{split}
\end{equation}

Equations $(\ref{eqn:sfft_eq16})$ and $(\ref{eqn:sfft_eq17})$ form a linear system with the array elements shown in Equation~$(\ref{eqn:sfft_eq18})$. 

\begin{equation}
\label{eqn:sfft_eq18}
\begin{tikzpicture}[
style1/.style={
  matrix of math nodes,
  every node/.append style={text width=#1,align=center,minimum height=2.5ex},
  nodes in empty cells,
  left delimiter=[,
  right delimiter=],
  },
style2/.style={
  matrix of math nodes,
  every node/.append style={text width=#1,align=center,minimum height=2.5ex},
  nodes in empty cells,
  left delimiter=\lbrace,
  right delimiter=\rbrace,
  }
]

\matrix[style1=0.3cm] (1mat)
{
  & & & & & \\
  & & & & & \\
  & & & & & \\
  & & & & & \\
  & & & & & \\
  & & & & & \\
  & & & & & \\
  & & & & & \\
  & & & & & \\
};

\draw[dashed]
  (1mat-6-1.south west) -- (1mat-6-6.south east);
\draw[dashed]
  (1mat-1-4.north east) -- (1mat-9-4.south east);

\node[font=\small] 
  at (1mat-3-2.south east) {$A_{\bar{i}\bar{j}\bar{\alpha}\bar{\beta} ij\alpha\beta}$};
\node[font=\small] 
  at ([xshift=3.5pt]1mat-3-5.south east) {$B_{\bar{i}\bar{j}\bar{\alpha}\bar{\beta} pq}$};
\node[font=\small]
  at (1mat-8-2.south east) {$\tilde{B}_{\bar{p}\bar{q} ij\alpha\beta}$};
\node[font=\small] 
  at ([xshift=3.5pt]1mat-8-5.south east) {$C_{\bar{p}\bar{q} pq}$};

\matrix[style2=0.05cm,right=20pt of 1mat] (2mat)
{
  & \\
  & \\
  & \\
  & \\
  & \\
  & \\
  & \\
  & \\
  & \\
};
\draw[dashed]
  (2mat-6-1.south west) -- (2mat-6-2.south east);

\node[font=\small]
  at (2mat-3-1.south east) {$a_{\bar{i}\bar{j}\bar{\alpha}\bar{\beta}}$};
\node[font=\small]
  at (2mat-8-1.south east) {$b_{pq}$};

\node at ([xshift=15pt,yshift=-1.2pt]2mat.east) {$=$};

\matrix[style2=0.05cm,right=30pt of 2mat] (3mat)
{
  & \\
  & \\
  & \\
  & \\
  & \\
  & \\
  & \\
  & \\
  & \\
};
\draw[dashed]
  (3mat-6-1.south west) -- (3mat-6-2.south east);
  
\node[font=\small]
  at (3mat-3-1.south east) {$D_{\bar{i}\bar{j}\bar{\alpha}\bar{\beta}}$};
\node[font=\small]
  at (3mat-8-1.south east) {$E_{\bar{p}\bar{q}}$};
\end{tikzpicture}
\end{equation}

The elements of Equation~(\ref{eqn:sfft_eq18}) are given explicitly as the following,

\begin{equation}
A_{\bar{i}\bar{j}\bar{\alpha}\bar{\beta} ij\alpha\beta} = \sum_{lm} \{\Re\;[\widehat{R^{\bar{i} \bar{j}}} \widehat{\mathcal{K}_{\bar{\alpha}  \bar{\beta}}} (\widehat{R^{ij}})^* (\widehat{\mathcal{K}_{\alpha \beta}})^*]\} (l, m),
\label{eqn:sfft_eq19}
\end{equation}

\begin{equation}
B_{\bar{i}\bar{j}\bar{\alpha}\bar{\beta} pq} = \sum_{lm} \{\Re\;[\widehat{R^{\bar{i} \bar{j}}} \widehat{\mathcal{K}_{\bar{\alpha}  \bar{\beta}}} (\widehat{T^{pq}})^*]\} (l, m),
\label{eqn:sfft_eq20}
\end{equation}

\begin{equation}
\tilde{B}_{\bar{p}\bar{q} ij\alpha\beta} = \sum_{lm} \{\Re\;[\widehat{T^{\bar{p}\bar{q}}} (\widehat{R^{ij}})^* (\widehat{\mathcal{K}_{\alpha\beta}})^*]\} (l, m),
\label{eqn:sfft_eq21}
\end{equation}

\begin{equation}
C_{\bar{p}\bar{q} pq} = \sum_{lm} \{\Re\;[\widehat{T^{\bar{p}\bar{q}}} (\widehat{T^{pq}})^*]\} (l, m),
\label{eqn:sfft_eq22}
\end{equation}

\begin{equation}
D_{\bar{i}\bar{j}\bar{\alpha}\bar{\beta}} = \sum_{lm} \{\Re\; [\widehat{S}^* \widehat{R^{\bar{i}\bar{j}}} \widehat{\mathcal{K}_{\bar{\alpha}  \bar{\beta}}}]\} (l, m),
\label{eqn:sfft_eq23}
\end{equation}

\begin{equation}
E_{\bar{p}\bar{q}} = \sum_{lm} \{\Re\;[\widehat{S}^* \widehat{T^{\bar{p}\bar{q}}}]\} (l, m).
\label{eqn:sfft_eq24}
\end{equation}

These equations can be further simplified as shown in Appendix~\ref{App:simplify}. Finally, the difference image $D$ is calculated from Equation~(\ref{eqn:sfft_eq13}) with applying the linear system solution $\{...,a_{ijab}, ..., b_{pq}, ...\}$ of Equation~(\ref{eqn:sfft_eq18}).
Given that the difference image will possess correlated noise on the background, we present a decorrelation process to whiten the background noise in Appendix~\ref{App::noise-decorr}.

\section{The implementation of SFFT} \label{sec:sfft_implement}

\begin{figure*}[ht!] 
    \centering
    \includegraphics[trim=1cm 0cm 1cm 0cm,clip=true,width=13cm]{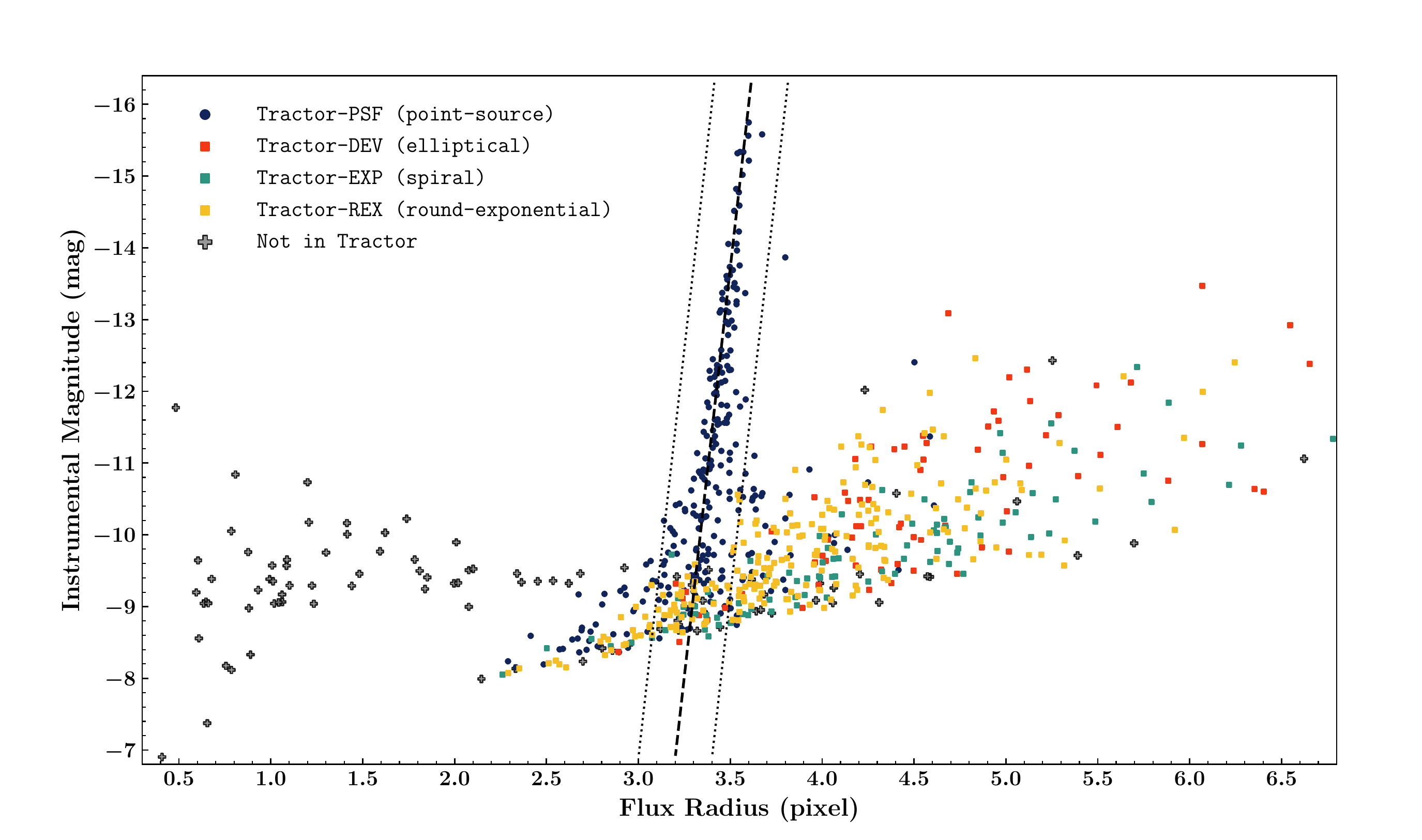}
    \caption{\label{fig:fig1} {Morphological classifier in $\textit{sparse-flavor}$ SFFT demonstrated by an arbitrary DECam image of Legacy Survey \citep{LegacySurvey}. The classifier is based on the photometric results of the image using {\tt\string SEXTRACTOR}. The data points show the relationship between instrumental magnitude ({\tt\string SEXTRACTOR} catalog value $\textrm{MAG\_AUTO}$) and flux radius ({\tt\string SEXTRACTOR} catalog value $\textrm{FLUX\_RADIUS}$) for the detected objects with {\tt\string SEXTRACTOR} catalog value $\textrm{FLAG}$ being zero. We also group the data by the morphological types in the Tractor catalog of Legacy Survey \citep{LegacySurvey} showing with different markers in the figure.
    The dashed black line is detected by Hough Transformation as the strongest straight line feature with vertical orientation. Two dotted lines are parallel to the dashed line with fixed separation being 0.2 (SFFT parameter $\textbf{\text{-BeltHW}}$), the region between which is referred to as $\textit{point-source-belt}$.}}
\end{figure*}

\begin{deluxetable*}{llllllll}[ht!]
    \tablenum{1}
    \tablewidth{0pt}
    \tablecolumns{6}
    \tablecaption{\label{tab:inst_spec} Technical Specification of the Instruments}
    
    \tablehead{
    \colhead{Instrument}               &
    \colhead{Telescope}                &
    \colhead{Pixel Scale}              &
    \colhead{Field of View}            &
    \colhead{Field Property}           &
    \colhead{Comments}
    \\
    \colhead{}                         &
    \colhead{(m)}                      &
    \colhead{(arcsec/pix)}             &
    \colhead{(deg$^2$)}                &
    \colhead{}                         &
    \colhead{}
    }
    
    \startdata
    $\quad\text{ZTF}$&      $\quad\,\,\text{1.2}$      &      $\quad\quad1.0$      &      $\quad\quad\text{47.0}$      &      $\quad\,\,\text{crowded}$      &      $\quad\quad\quad\text{M31}$
    \\
    $\quad\text{AST3-II}$&      $\quad\,\,\text{0.5}$      &      $\quad\quad1.0$      &      $\quad\quad\text{4.3}$      &      $\quad\,\,\text{crowded}$      &      $\quad\quad\quad\text{LMC}$
    \\
    $\quad\text{TESS}$&      $\quad\,\,4\times0.1$      &      $\quad\quad21.0$      &      $\quad\quad4\times576.0$      &      $\quad\,\,\text{crowded}$      &      $\quad\quad\quad\text{space-based}$
    \\
    $\quad\text{DECam}$&      $\quad\,\,\text{4.0}$      &      $\quad\quad0.262$      &      $\quad\quad\text{3.0}$      &      $\quad\,\,\text{sparse}$      &      $\quad\quad\quad\text{-}$
    \\
    $\quad\text{TMTS}$&      $\quad\,\,4\times0.4$      &      $\quad\quad1.86$      &      $\quad\quad4\times4.5$      &      $\quad\,\,\text{sparse}$      &      $\quad\quad\quad\text{CMOS detectors}$
    \enddata
\end{deluxetable*}

We have presented the mathematical derivation of the SFFT image difference algorithm in Section~\ref{subsec:m-subtract}. In a nutshell, to calculate the difference image $D$ from a pair of the reference image ($R$) and science image ($S$), one can first search for the optimal parameters $\{..., a_{ijab}, ..., b_{pq}, ...\}$ by minimizing the power of the resulting difference image of $R$ and $S$ in Fourier space via a linear system described in Equation~(\ref{eqn:sfft_eq18}). Subsequently, one needs to apply the solution to match image $R$ to image $S$ following Equation~(\ref{eqn:sfft_eq13}) and get the ultimate difference image $D$.

However, saturated sources with a significant non-linear response, casual cosmic rays and moving objects, bad CCD pixels, optical ghosts, and even the variable objects and transients themselves can severely affect the construction of reliable convolution kernels. These objects can be seen as distractions for image subtraction (hereafter referred to as $\textit{distraction-sources}$), which are hardly modeled by the image subtraction algorithm but ubiquitous in real observations.
We need further to offer an effective channel to prevent the $\textit{distraction-sources}$ in $R$ and $S$ from contributing to the parameter-solving process.
One likely choice is to deweight those trivial pixels, but it appears to be tricky to formulate the weight assignment in Fourier space. Instead, we found that a more straightforward approach by adequate image masking is easier to implement and can perform well enough in our study.

In the current implementation of SFFT, image subtraction is developed as a two-step process. First, make a masked version of reference image $R$ and science image $S$, denoted as $\breve{R}$ and $\breve{S}$, respectively. Second, solve the linear system in Equation~(\ref{eqn:sfft_eq18}) using the masked image-pair $\breve{R}$ and $\breve{S}$, then apply the solution to calculate the difference image $D$ of the original image-pair $R$ and $S$ following Equation~(\ref{eqn:sfft_eq13}).
The SFFT algorithm does not rely on isolated stellar objects to construct PSF-matching kernels, thereby performing equally well for the observations taken in crowded and sparse fields. 
Our software provides two flavors for image subtraction, $\textit{crowded-flavor}$ and $\textit{sparse-flavor}$, to accommodate the situations for both crowded and sparse fields. But the flavor for crowded fields can also work for sparse fields, only that the results may be affected by the large number of noisy background pixels that do not contribute to the construction of the convolution kernel. 
The two flavors follow the same routine for image subtraction and differ only in ways of masking the data. 

\subsection{Preprocessing for Crowded-Flavor SFFT} \label{subsec:preproc-crowded}

Applying SFFT is straightforward for signal-dominated crowded fields. 
Mostly, saturation is the predominant factor in all $\textit{distraction-sources}$ for crowded fields. The $\textit{crowded-flavor}$ SFFT will automatically mask the pixels affected by the saturated sources in the field and replace them with the neighboring background. To further reduce image matching errors, it is allowed to use a more elaborate customized mask instead, e.g., to include the pixels affected by known variables and transients in the field. 

\subsection{Preprocessing for Sparse-Flavor SFFT} \label{subsec:preproc-sparse}

On the flip side, the $\textit{sparse-flavor}$ SFFT is relatively more sophisticated on image masking. We select a set of sources of astronomical significance in the field and mask all other irrelevant regions so that the selected objects can dominate the parameters-solving process. It is reminiscent of the similar strategy in {\tt\string HOTPANTS}, which fits on the rectangle stamps (i.e., ``sub-stamps" in {\tt\string HOTPANTS} terminology) encompassing selected astronomical objects.
The analogous intention of both methods is to restrict the calculation on a properly pre-selected set of sources to eliminate the effect of $\textit{distraction-sources}$ on the solution of convolution kernels.
In this scheme, the regions masked by SFFT have also included abundant background pixels. This is a reasonable consideration in our framework. Pixel uncertainty is not considered as weights in SFFT subtraction. Accordingly, the overwhelming noise-dominated background pixels are more likely to degrade the construction of accurate PSF-matching kernels rather than to contribute (see the same consideration in \citet{Kochanski96}). 

In practice, the input image-pair of $\textit{sparse-flavor}$ SFFT is required to be sky subtracted. This requirement is to simplify the image-masking process so that all the pixels enclosed in masked regions can be replaced by a constant of zero. As a result, the differential background term in the SFFT algorithm becomes trivial as the background offset between the input image-pair has been minimized by sky subtraction. That is to say, we pass on the function of matching differential background, which was embedded in numerical calculations of SFFT, to a customized sky subtraction as a preliminary process. Fortunately, modeling sky background is usually feasible in sparse fields (e.g., using an interpolation-based method). Note this requirement is not a prerequisite for $\textit{crowded-flavor}$ SFFT. This is because properly modeling sky background can be tricky for crowded fields: the modeled sky is more susceptible to being biased by the signal harboring in the pixels misidentified as background (e.g., the outskirts of nearby galaxies, see \citet{NoiseChisel}).

We developed a morphological classifier to carry out the source selection in $\textit{sparse-flavor}$ SFFT. The classifier was initially proposed for {\tt\string PSFEx} \citep{PSFEx}, which enables {\tt\string PSFEx} to select a subset of point sources for constructing the PSF model. 
Given a photometry catalog generated by {\tt\string SEXTRACTOR} \citep{SExtractor}, one can draw a figure of instrumental magnitudes ({\tt\string SEXTRACTOR} catalog value $\textrm{MAG\_AUTO}$) against flux radius ({\tt\string SEXTRACTOR} catalog value $\textrm{FLUX\_RADIUS}$) for all detected photometry objects. 
Generally, bright point sources tend to stay around a nearly vertical straight line on this figure. {\tt\string PSFEx} leverages the statistical feature to select appropriate samples to establish the PSF model. 
Although the goal of source selection for $\textit{sparse-flavor}$ SFFT is not entirely aligned to that for {\tt\string PSFEx}, it inspires us to make the selection based on {\tt\string SEXTRACTOR} parameters and their statistical characters.

To demonstrate the source selection criterion in $\textit{sparse-flavor}$ SFFT, we show an example of an individual DECam \citep{DECam2008arXiv0810.3600H} image in Figure~\ref{fig:fig1}. By contrast, the {\tt\string SEXTRACTOR} photometry catalog of this image is cross-matched with the Tractor catalog from Legacy Survey \citep{LegacySurvey}, which offers their own fitted morphological types. 
As shown in Figure~\ref{fig:fig1}, the point sources and extended sources form two conspicuous branches, which intersect at the faint end but become well-separated towards the bright side. 
According to the Tractor types, the branch with a nearly vertical orientation primarily comprises point-like sources with Tractor type PSF, while the extended sources with Tractor types DEV, EXP, REX are likely to be found in the other branch that sprawls out horizontally. Moreover, the discrete gray dots represent the objects that cannot be found in the Tractor catalog, which are casual detections such as cosmic rays. 
SFFT uses Hough Transformation \citep{Hough:1959qva} to identify the straight line surrounded by the branch of point sources.  
The figure is first pixelized by counting the number of objects in each small grid. Hough Transformation is applied so that we can search for the strongest line feature with vertical orientation. The thin belt-like region around the detected straight line with a fixed width is referred to as $\textit{point-source-belt}$. In SFFT, the parameter $\textbf{\text{-BeltHW}}$ describes the belt half-width (default value is 0.2). With this terminology, it is easier to describe the specific selection criteria.

For the source selection, $\textit{sparse-flavor}$ SFFT will first generate a {\tt\string SEXTRACTOR} photometry catalog for reference image and science image, respectively. A source is selected if it lies in $\textit{point-source-belt}$ or out of $\textit{point-source-belt}$ from the right side for both images. The extended sources in the field are not excluded, as SFFT can use both stars and galaxies to derive the solution for the matching kernel. 
However, typical cosmic rays and the faintest astronomical sources will be discarded. Given that each pixel of selected samples is equally weighted no matter bright or faint in SFFT, rejecting the noisiest subset should be appropriate in our framework. 
Furthermore, {\tt\string SEXTRACTOR} has been configured to only output the objects with {\tt\string SEXTRACTOR} catalog value $\textrm{FLAG}$ being zero. Namely, this guarantees that the saturated sources and blending objects (relatively rare in sparse fields) are not in the selection.

Significant variable sources in the field should be further rejected. SFFT cross-matches the photometry catalogs of the input image-pair, then calculates the difference in instrumental magnitude for each matched object. The outliers of the distribution of magnitude difference indicating violent brightness change will be discarded from the initial set of selection. In SFFT, the outliers are identified by the threshold parameter $\textbf{\text{-MAGD$\_$THRESH}}$ (default value is 0.12 mag).
The survived selection set is applied for the image-masking in $\textit{sparse-flavor}$ SFFT. The outlier rejection is usually sufficient enough for a successful image subtraction, but one is also allowed to further refine the automatic selection by removing the known but not recognized variables and transients.
Unlike {\tt\string HOTPANTS}, SFFT does not enclose each source of the selection within a fixed-size rectangle stamp. Instead, we employ the detection mask of {\tt\string SEXTRACTOR} (check-image $\textrm{SEGMENTATION}$ generated by {\tt\string SEXTRACTOR}) to define the pixel domain of each selected source. Finally, the pixels not in any pixel domain will be masked by a constant of zero. 

\subsection{Implementation of SFFT Subtraction} \label{subsec:implement-sub}

\begin{figure}[t!]
    \centering
    \includegraphics[trim=0cm 0cm 0cm 0cm,clip=true,width=8.5cm]{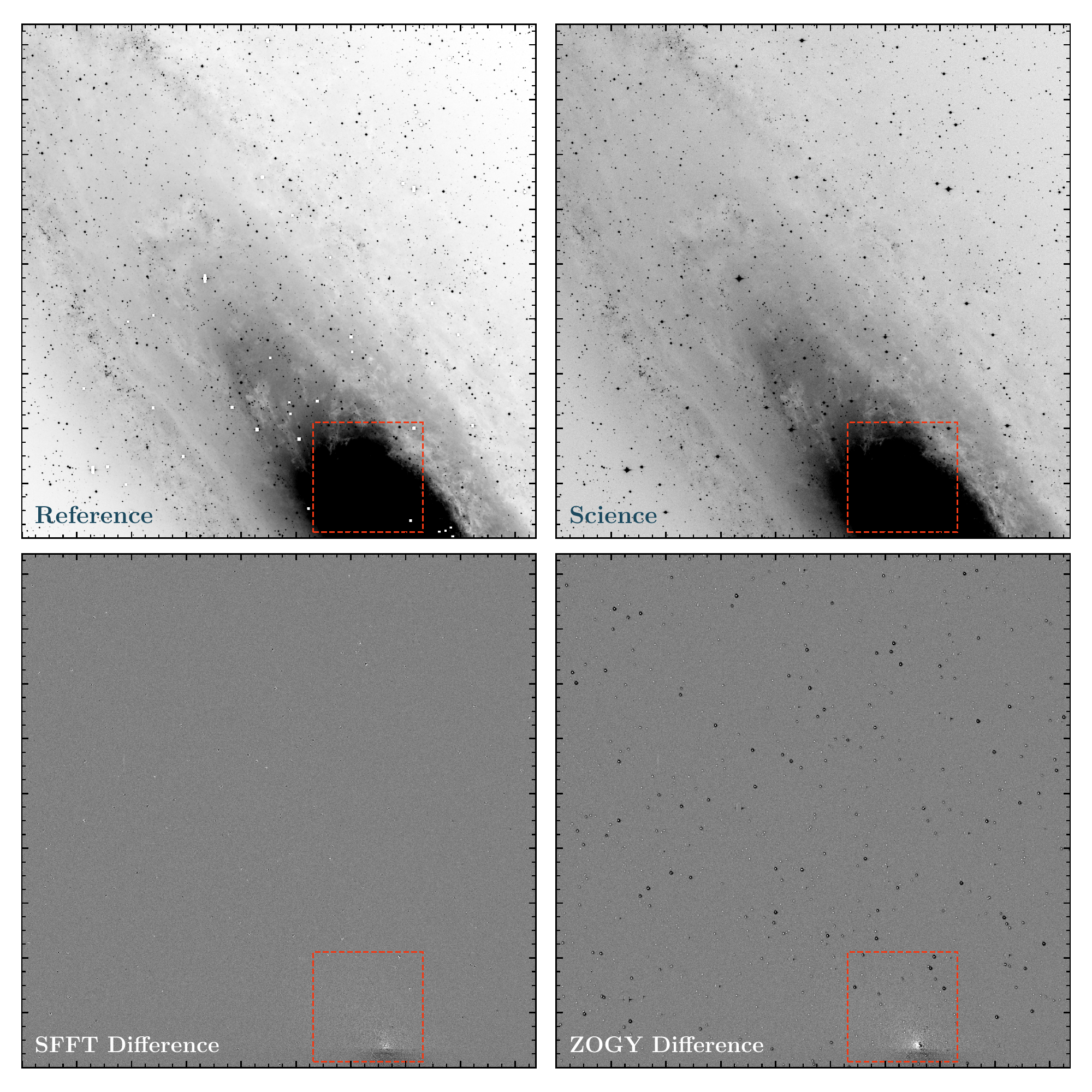}
    \caption{\label{fig:ZTF_Wide} {Image subtraction performance of the ZTF test with observations covering M31. The upper panel, from left to right, shows the input reference image \textit{ZTF-REF} and science image \textit{ZTF-SCI}. The lower panel presents the resulting difference image from SFFT (left) and ZOGY (right), where the ZOGY difference is directly downloaded from ZTF Data Release 3. The red dashed square marks the core region of the M31.}}
\end{figure}

\begin{figure}[t!]
    \centering
    \includegraphics[trim=0cm 0cm 0cm 0cm,clip=true,width=8.5cm]{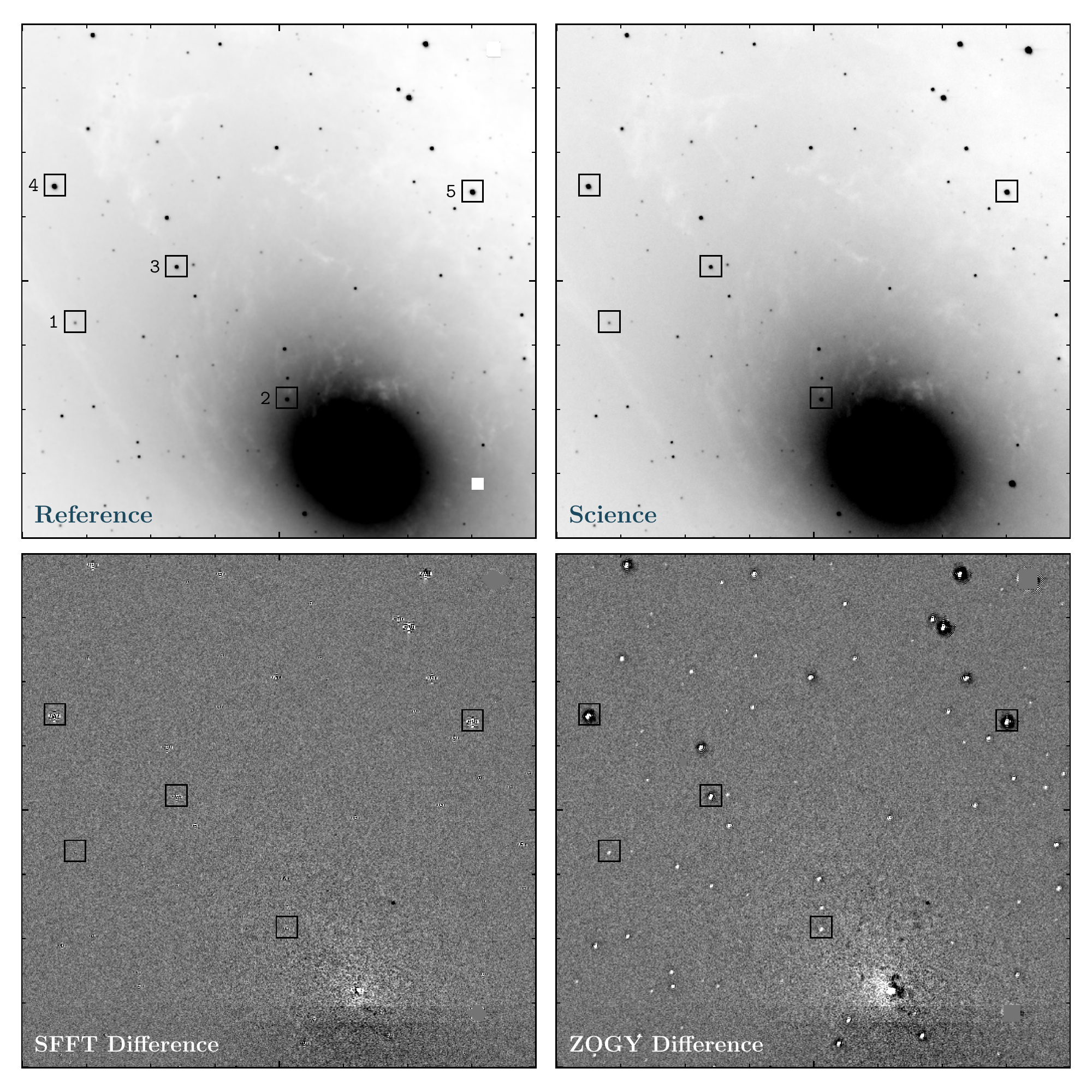}
    \caption{\label{fig:ZTF_Narrow} {Image subtraction performance of the ZTF test around M31 galaxy center. The panel descriptions are the same as Figure~\ref{fig:ZTF_Wide}, however, in a close-up view of the region enclosed with the red square in Figure~\ref{fig:ZTF_Wide}. In each panel, the solid black squares mark five field stars to be checked in detail in Figure~\ref{fig:ZTF_Bias}.}}
\end{figure}

\begin{figure}[ht!]
    \centering
    \includegraphics[trim=0.7cm 0.6cm 0.7cm 0.6cm,clip=true,width=8.5cm]{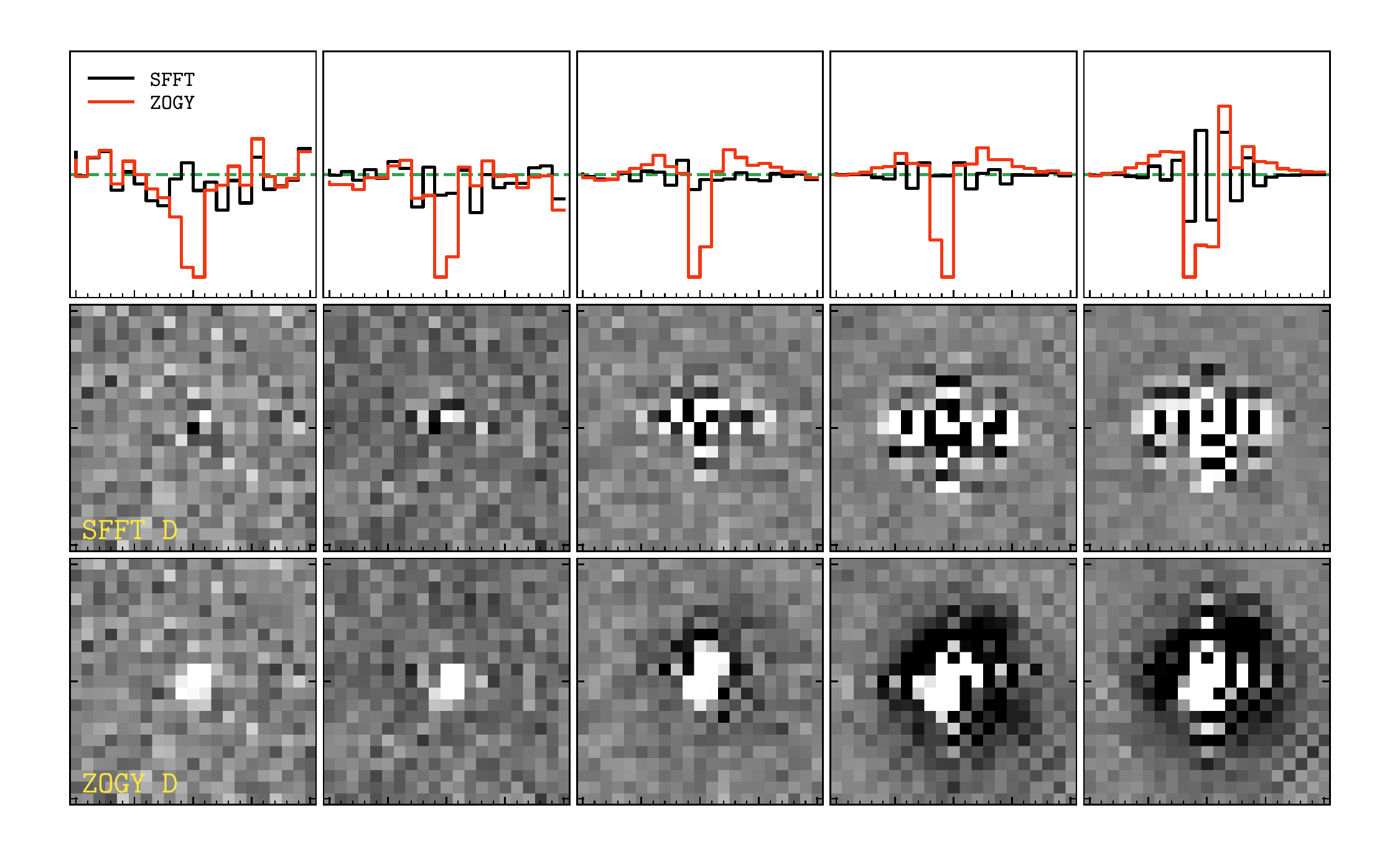}
    \caption{\label{fig:ZTF_Bias} {Analysis of flux residuals for the ZTF test. The second row and third row show the thumbnail images of the selected five samples from the SFFT difference and ZOGY difference, respectively. From left to right, thumbnail images are sorted by the sample labels shown in Figure~\ref{fig:ZTF_Narrow}. For each thumbnail image, flux integral along the y-axis can produce a flux curve as a function of the index of the x-axis. These functions are shown in the first row, using red solid curves for ZOGY and black solid curves for SFFT. In each panel of the first row, the two flux curves have been rescaled with the same factor for display clarity. The green dashed line shows the flux level of constant zero.}}
\end{figure}

\begin{figure*}[ht!]
    \centering
    \includegraphics[trim=0cm 1cm 0cm 1cm,clip=true,width=13cm]{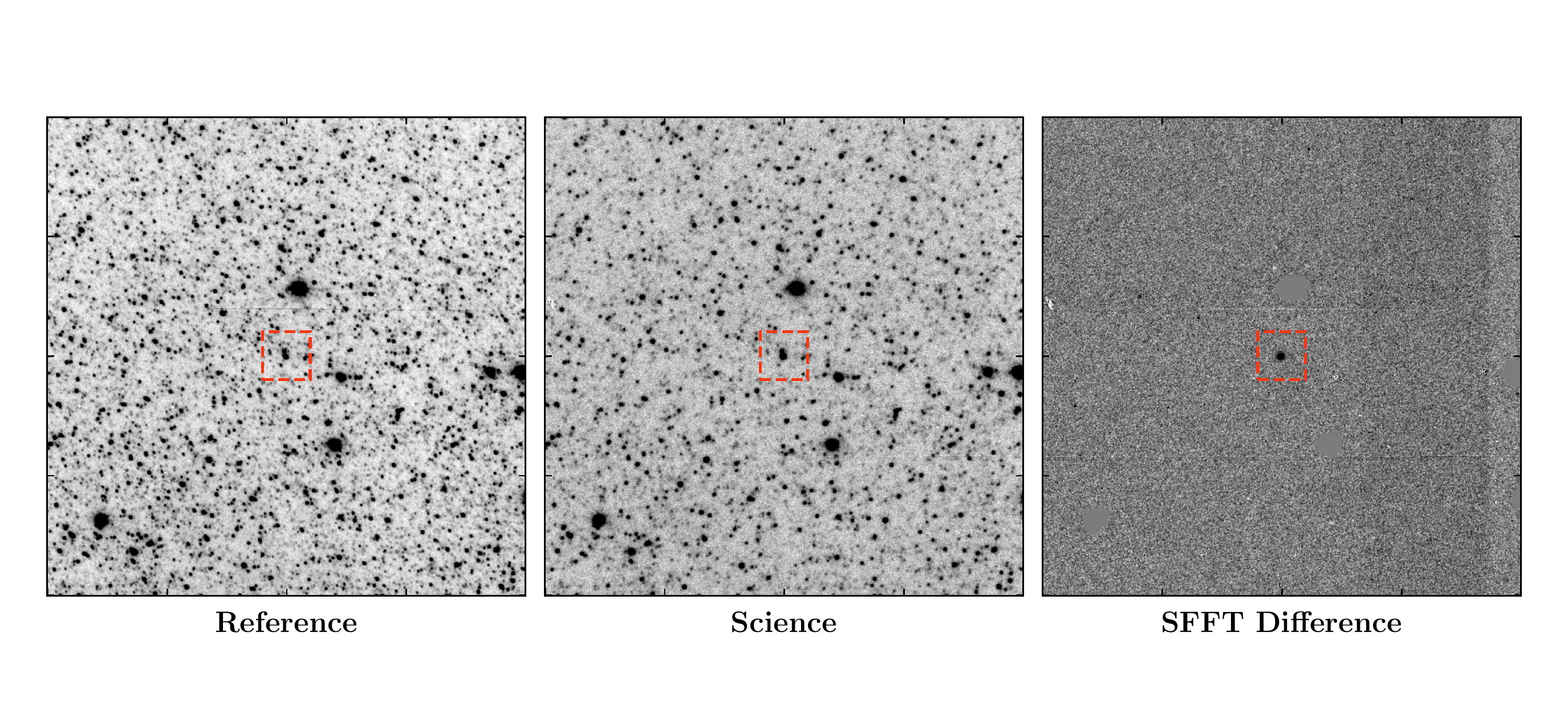}
    \caption{\label{fig:LMC_example} {Image subtraction performance of the AST3-II test with observations of LMC center. The figure, from left to right, shows the input reference image \textit{AST-REF}, science image \textit{AST-SCI} and resulting difference image by SFFT. The red dashed square at the image center indicates a known RRAB variable TY Dor (RA=05:24:06.35 DEC=-69:25:11.0) from VSX catalog \citep{VSX}. The vertical strip reveals the detection readout sections with slightly different gain values.}}
\end{figure*}

\begin{figure*}[ht!]
    \centering
    \includegraphics[trim=0cm 1cm 0cm 1cm,clip=true,width=13cm]{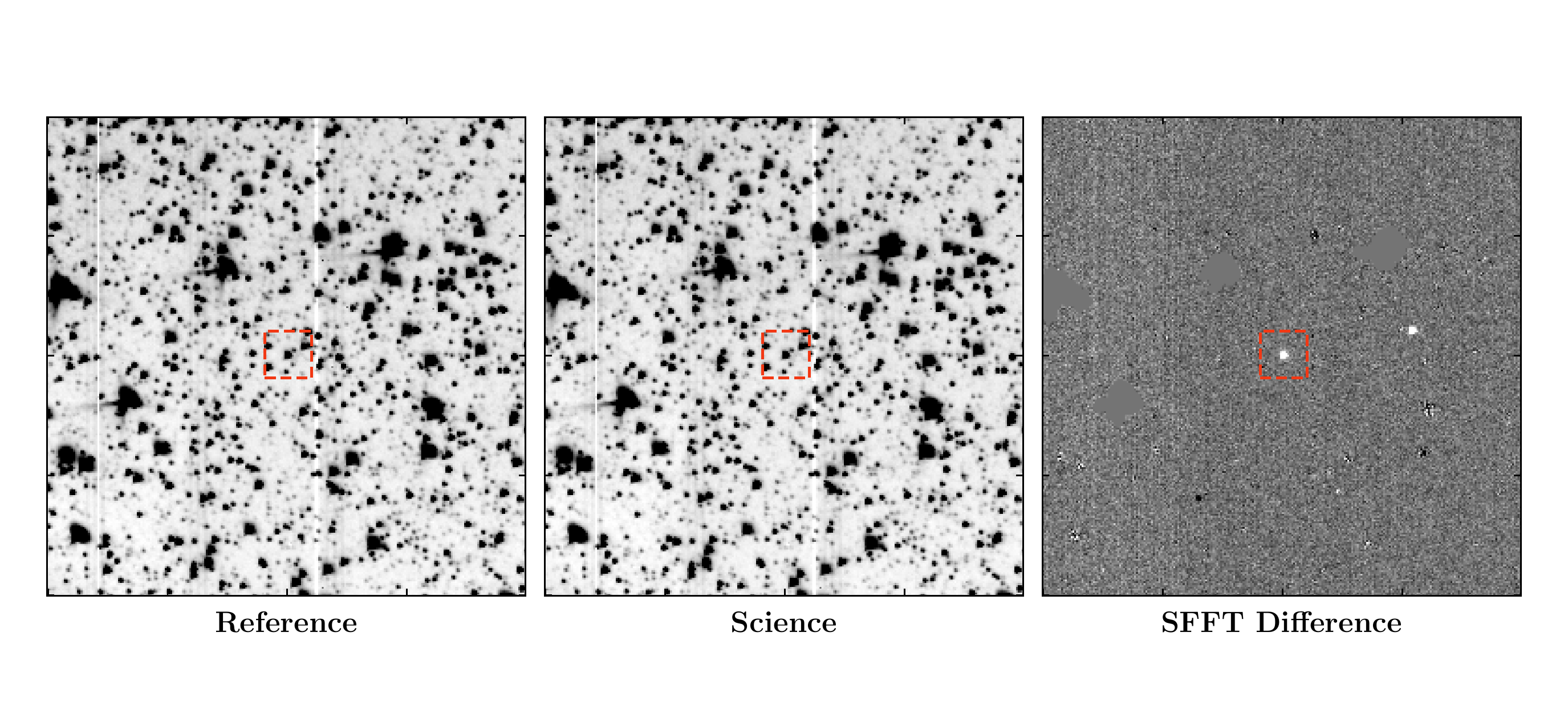}
    \caption{\label{fig:TESS_example} {Image subtraction performance of the TESS test. The figure, from left to right, shows the input reference image \textit{TESS-REF}, science image \textit{TESS-SCI} and resulting difference image by SFFT. The red dashed square at the image center indicates a known RRAB variable CV Scl (RA=23:09:30.52 DEC=-35:47:16.9) from VSX catalog \citep{VSX}.}}
\end{figure*}

The next step is to perform subtraction with the unmasked and the masked image-pairs, in the same way for the two flavors.
We implemented the subtraction algorithm in CUDA following the mathematical reasoning in Section~\ref{subsec:m-subtract}.

There are a few free parameters in this process. In Section~\ref{subsec:m-subtract}, the spatial variation of the convolution kernel was modeled as a polynomial form of degree $D_K$. Similarly, the differential background is assumed to be a polynomial with degree $D_B$. In SFFT, the two free parameters correspond to $\textbf{\text{-KerPolyOrder}}$ and $\textbf{\text{-BGPolyOrder}}$, respectively. The default values of the two parameters are 2.
Another parameter that should be specified is the pixel size of the convolution kernels. The default way is to determine kernel size as seeing-related, i.e., the kernel half-width is calculated as $\textbf{\text{-KerHWRatio}}\times\text{FWHM}_\text{L}$, where the ratio $\textbf{\text{-KerHWRatio}}$ is a free parameter of SFFT and $\text{FWHM}_\text{L}$ is the worse seeing of the input image-pair.
The last free parameter $\textbf{\text{-ConstPhotRatio}}$ is boolean, which controls whether SFFT subtraction is subject to a constant photometric ratio or not. The default value of this parameter is $\texttt{True}$ that assumes the input image-pair has been well-calibrated. Setting the parameter to be $\texttt{False}$ means the photometric ratio will also become a polynomial form of degree $\textbf{\text{-KerPolyOrder}}$.

\section{Examples of Image Subtraction with SFFT} \label{sec:sfft_examples}

We show examples of SFFT applied to observations from five different telescopes. These data correspond to a diverse range of characteristics. The technical specifications of the instruments are given in Table~\ref{tab:inst_spec}.
To test the two flavors of SFFT, we use ZTF \citep{ZTF_Eric18}, AST3-II \citep{AST3:2014SPIE.9145E..0FY}, and TESS \citep{TESS_Ricker15} data as examples of crowded fields. Also, the observations from DECam and TMTS with abundant isolated stars are selected as representatives of sparse fields. 

The ZTF \citep{ZTF_Bellm19} and AST3-II \citep{AST3-WangLingZhi2017AJ....153..104W,AST3:2014SPIE.9145E..0FY} data are images of the nearby galaxies M31 and the Large Magellanic Clouds (LMC), respectively. The two wide-field survey telescopes have the same pixel sampling but point to a different hemisphere. 
We additionally test SFFT with TESS \citep{TESS_Ricker15} images with a vastly different pixel scale of 21.0 arcsec/pixel where PSF is severely under-sampled. All of these selected observations could be seen as signal-dominated cases so that subtraction tests on them are conducted by the $\textit{crowded-flavor}$ SFFT.

The DECam and TMTS data were acquired for extragalactic transient surveys \citep{DECamERON_Mould17,TMTS_Zhang20}. The DECam data are from deep surveys with a large number of isolated point sources and extended galaxies, while the TMTS data are from a shallow nearby survey that recorded plenty of point sources. TMTS has a larger pixel scale than DECam, but its PSF is not under-sampled, unlike TESS. 
For these observations, the observed stellar objects are generally isolatedly distributed in the field, making it possible to apply the source selection described in Section~\ref{sec:sfft_implement}. We use $\textit{sparse-flavor}$ SFFT for the subtraction tests on DECam and TMTS.

Through this section, SFFT always uses the default configuration of free parameters described in Section~\ref{sec:sfft_implement}, which are summarized in Table~\ref{tab:sfft_param}, even though the two sets of images differ drastically.
The detailed information for the test data used in this section is presented in Table~\ref{tab:test_data_info}. In the context, we will highlight the alias names of the test images in italic.
All input image-pairs for image subtraction have been registrated by {\tt\string SWarp} \citep{SWarp} with LANCZOS3 resampling function. For the input images of $\textit{sparse-flavor}$ SFFT, we have modeled and subtracted their sky using {\tt\string SEXTRACTOR} with a mesh size of 64 $\times$ 64 pixels.

\begin{figure}[t!]
    \centering
    \includegraphics[trim=0cm 0cm 0cm 0cm,clip=true,width=8.5cm]{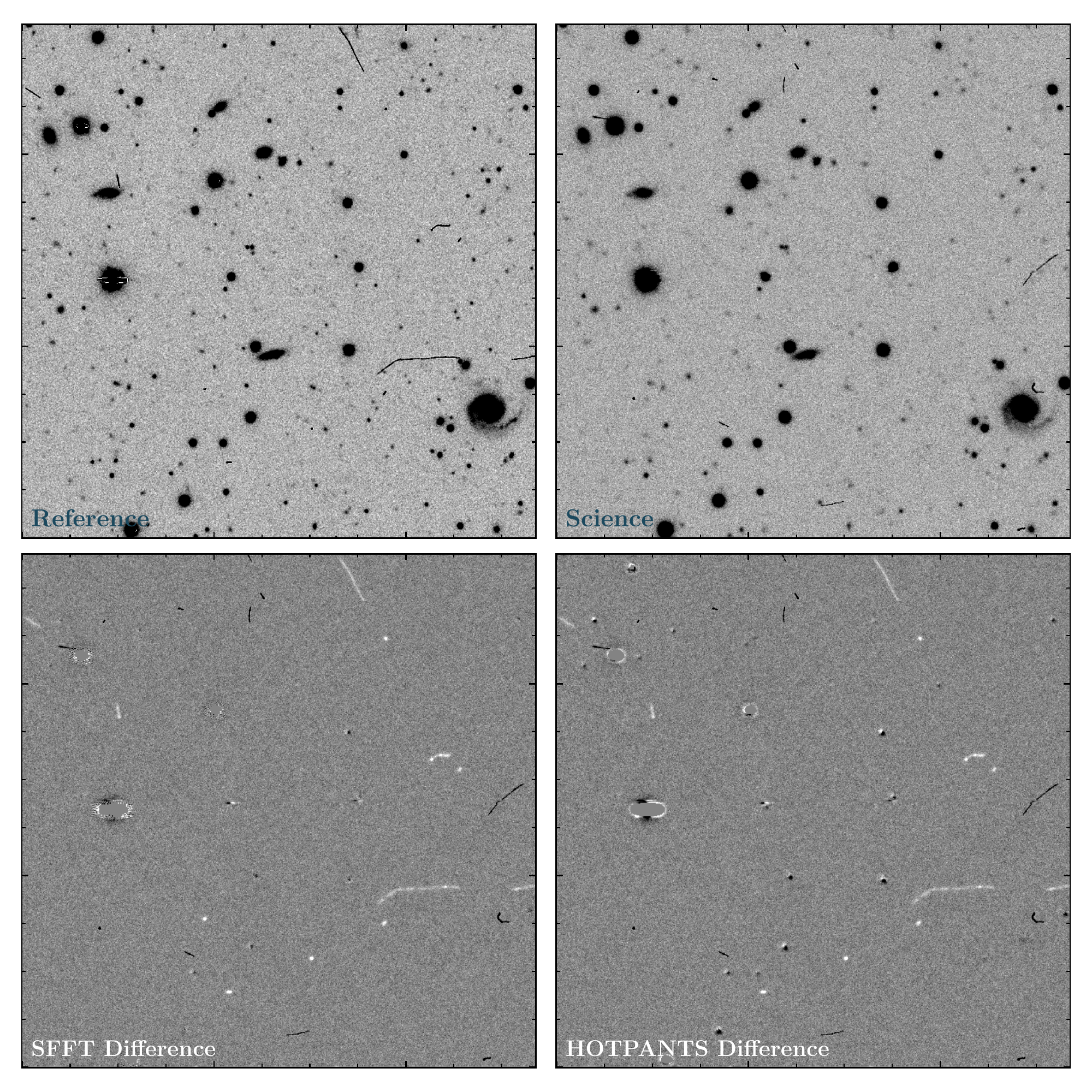}
    \caption{\label{fig:DECam_example} {Image subtraction performance of the DECam test from DECamERON survey. The upper panel shows the input reference image $\textit{DECam-SREF}$ and science image $\textit{DECam-OBS02f}$. The lower panel shows the resulting difference image by SFFT (left) and {\tt\string HOTPANTS} (right). The full field of view of DECam is too large for display, here we only present them in a narrow view of 3 arcmin width.}}
\end{figure}

\begin{figure}[t!]
    \centering
    \includegraphics[trim=0cm 0cm 0cm 0cm,clip=true,width=8.5cm]{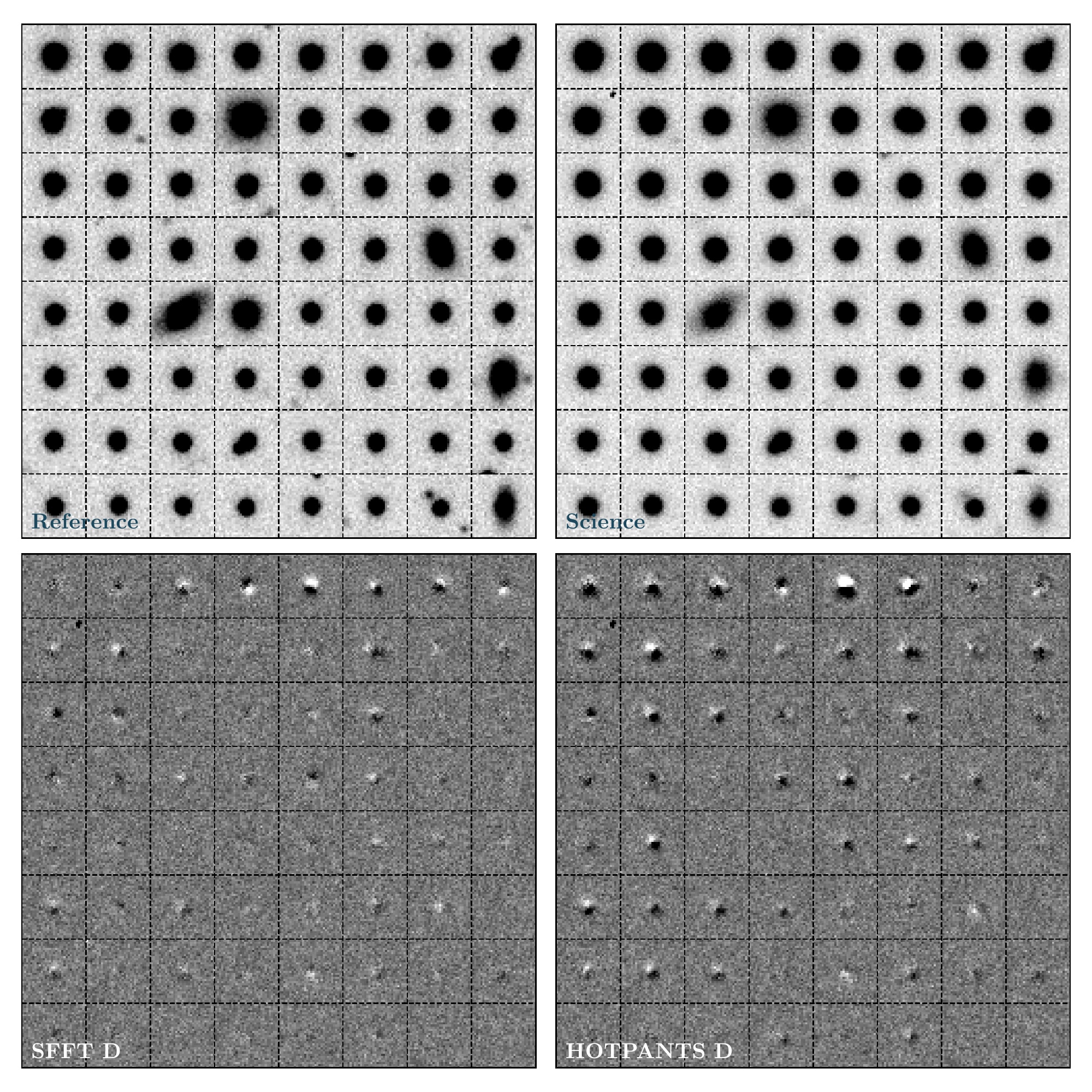}
    \caption{\label{fig:DECam_sample} {Image subtraction performance of the DECam test on the selected sources. The panels are arranged as in Figure~\ref{fig:DECam_example} but showing the subtraction results on the set of pre-selected individual sources. Each panel shows a synthetic image that combines thumbnail cutouts of 64 sources from the original images. The presented samples are drawn from the brightest end of the complete selection, and the thumbnails in the four panels are placed in order of decreasing brightness.}}
\end{figure}

\subsection{Crowded Fields} \label{subsec:crowded-tests}

By design, SFFT does not require isolated sources to solve the image matching. In crowded fields where few or no isolated stars are present, SFFT can perform exceptionally well.
The ZTF test images are downloaded from ZTF Data Release 3 \footnote{\url{https://www.ztf.caltech.edu/page/dr3}}, covering the nearby galaxy M31. Figure~\ref{fig:ZTF_Wide} and Figure~\ref{fig:ZTF_Narrow} show the subtraction performance of both the SFFT and ZOGY method, with a wider view and a close-up view on the galaxy core, respectively.

The difference image generated by SFFT is much cleaner than the ZOGY approach, as indicated by the lower panels of the two figures.
One can also notice that the difference image by the ZOGY method appears to have a bias seemingly due to the photometric mismatch of image subtraction. This guess is confirmed in Figure~\ref{fig:ZTF_Bias} that analyses the flux residuals of five field stars near the galaxy core. ZOGY subtraction uniformly remains negative net flux for these samples, suggesting an improperly-estimated photometric ratio between reference and science image. By contrast, SFFT does not have the same problem. 
Unlike ZOGY, the photometric ratio (a constant in our default configuration, i.e., $\mathring{a}_{0000}$) in SFFT is straightforwardly solved from the linear system, rather than an extra value calculated by some other photometric process.

Figure~\ref{fig:LMC_example} and Figure~\ref{fig:TESS_example} show SFFT subtraction performance on LMC observations of the Antarctica telescope AST3-II and test data from space-based TESS 30min-cadence Full Frame Images (FFIs), respectively.
A known variable star of RRAB type is easily identified at the image center for both cases. It illustrates how the flux variability can be effectively captured by SFFT subtraction from the densely packed fields.

\subsection{Sparse Fields} \label{subsec:sparse-tests}

\begin{figure}[t!]
    \centering
    \includegraphics[trim=0cm 0cm 1cm 0cm,clip=true,width=8.5cm]{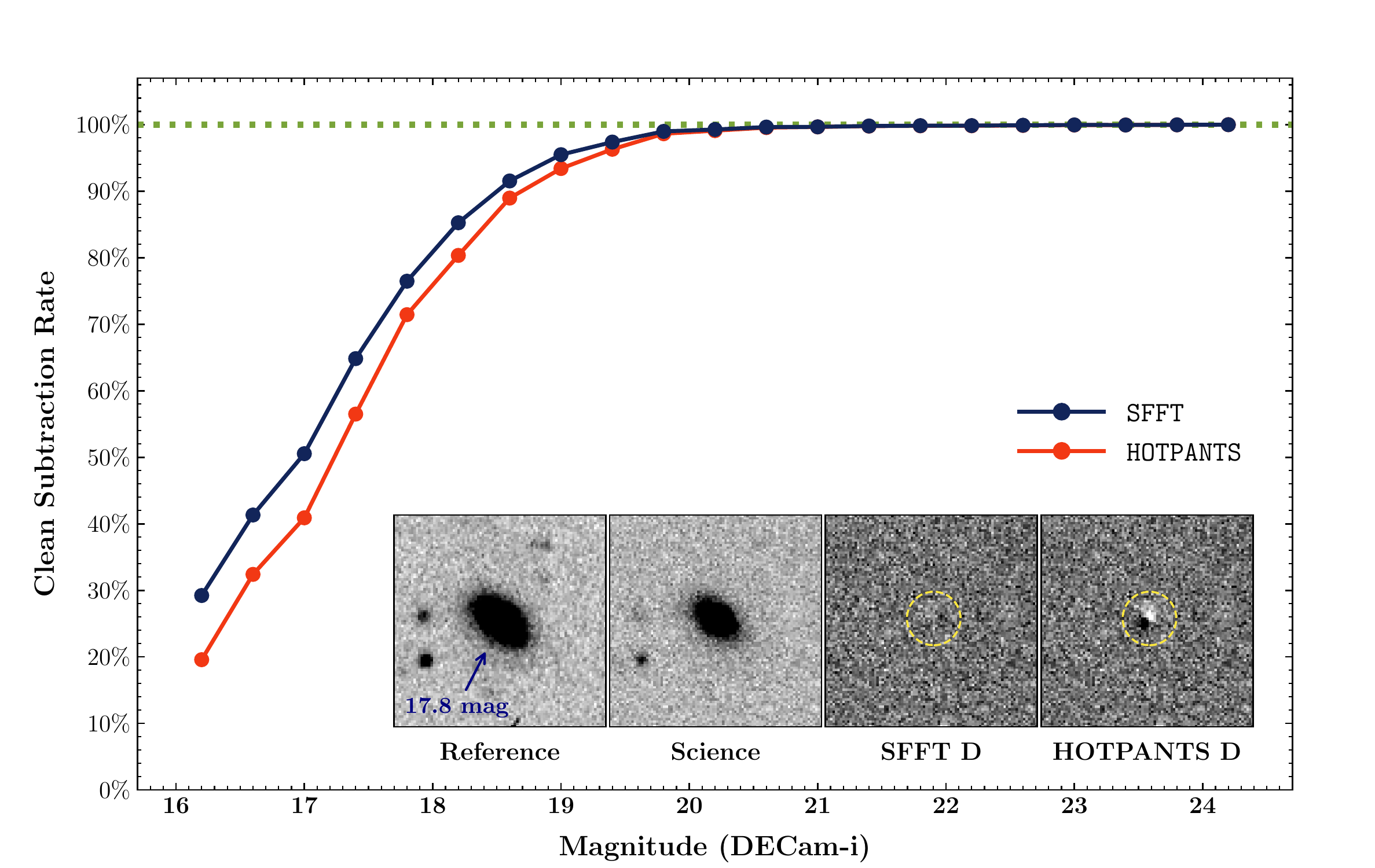}
    \caption{\label{fig:GalSub} {Clean subtraction rate of galaxies measured from the DECam test. The solid curves show clean subtraction rate as a function of galaxy magnitude for SFFT (blue) and {\tt\string HOTPANTS} (red), with magnitude bin width being 0.4 mag. The inset panels show an example of the subtraction performance for a bright galaxy. The first two panels of the insert show the zoomed-in region around the galaxy from the reference image $\textit{DECam-SREF}$ and the science image $\textit{DECam-OBS18a}$, respectively. The last two panels give the resulting difference image by SFFT and {\tt\string HOTPANTS}. The yellow dashed circle indicates the searching region for subtraction-induced artifacts with a radius of 11 pixels. The example represents the typical cases when SFFT is able to provide a clean subtraction but {\tt\string HOTPANTS} failed.}}
\end{figure}

\begin{figure}[t!]
    \centering
    \includegraphics[trim=0cm 0cm 1cm 0cm,clip=true,width=8.5cm]{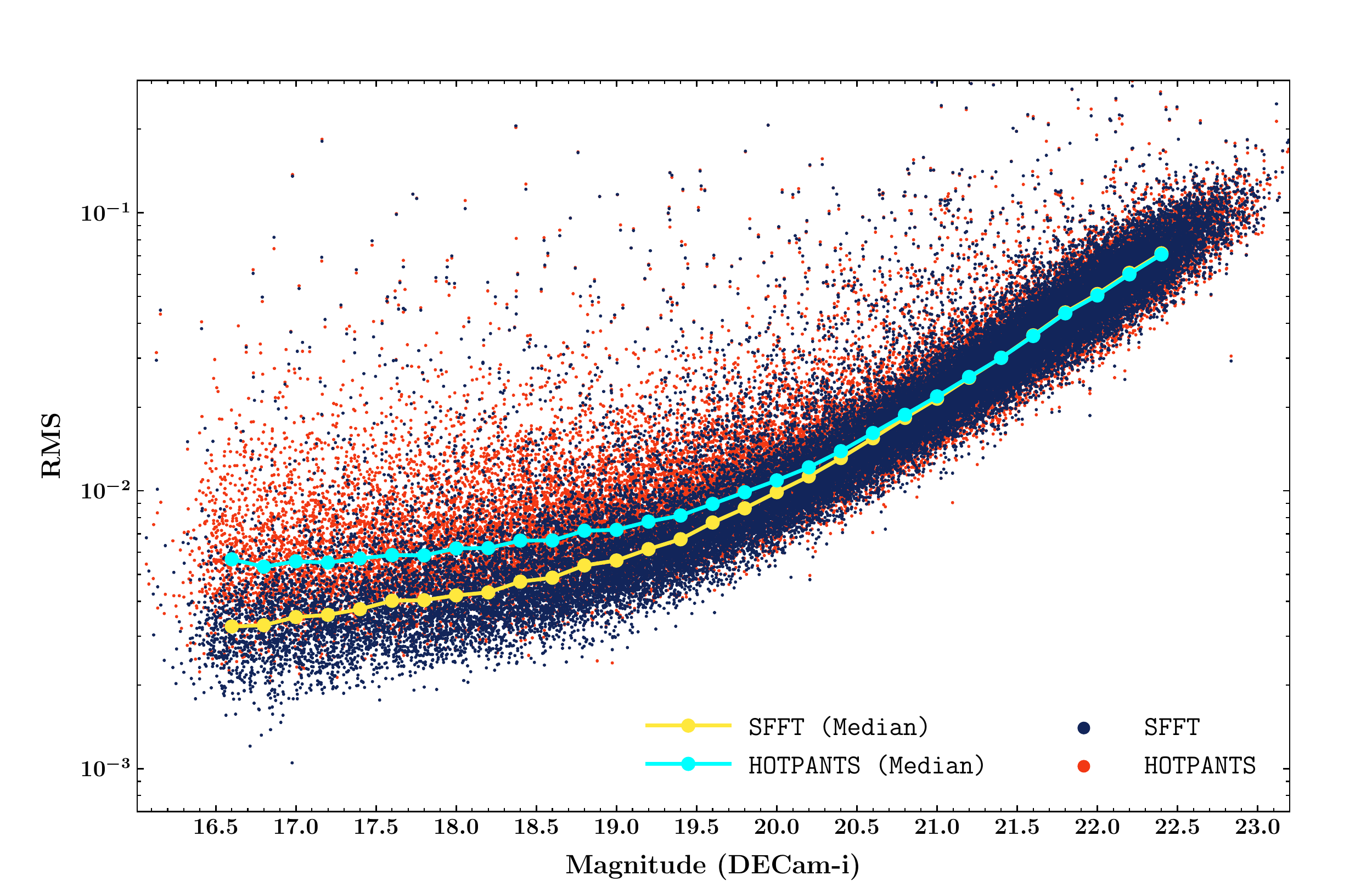}
    \caption{\label{fig:PhotAcc} Photometric accuracy measured from lightcurves of field stars for DECam test. The light curve of each field star (classied as PSF in Legacy survey) is derived by force aperture photometry on the queue of the convolved images (aliased with initial $\textit{DECam-OBS}$). The force photometry is carried out by {\tt\string SEXTRACTOR} with Gaussian optimal aperture $1.3462 \times \text{FWHM}_{\text{PTAR}}$, where $\text{FWHM}_{\text{PTAR}}$ is the FWHM of the specific CCD tile of $\textit{DECam-PTAR}$. The photometric uncertainty against star magnitude is shown as blue dots (for SFFT) and red dots (for {\tt\string HOTPANTS}). The solid curves give the median level of RMS in each magnitude bin with width 0.2 mag for SFFT (yellow) and {\tt\string HOTPANTS} (cyan), respectively.}
\end{figure}

\begin{figure*}[ht!]
    \centering
    \includegraphics[trim=0cm 1cm 0cm 0cm,clip=true,width=16.5cm]{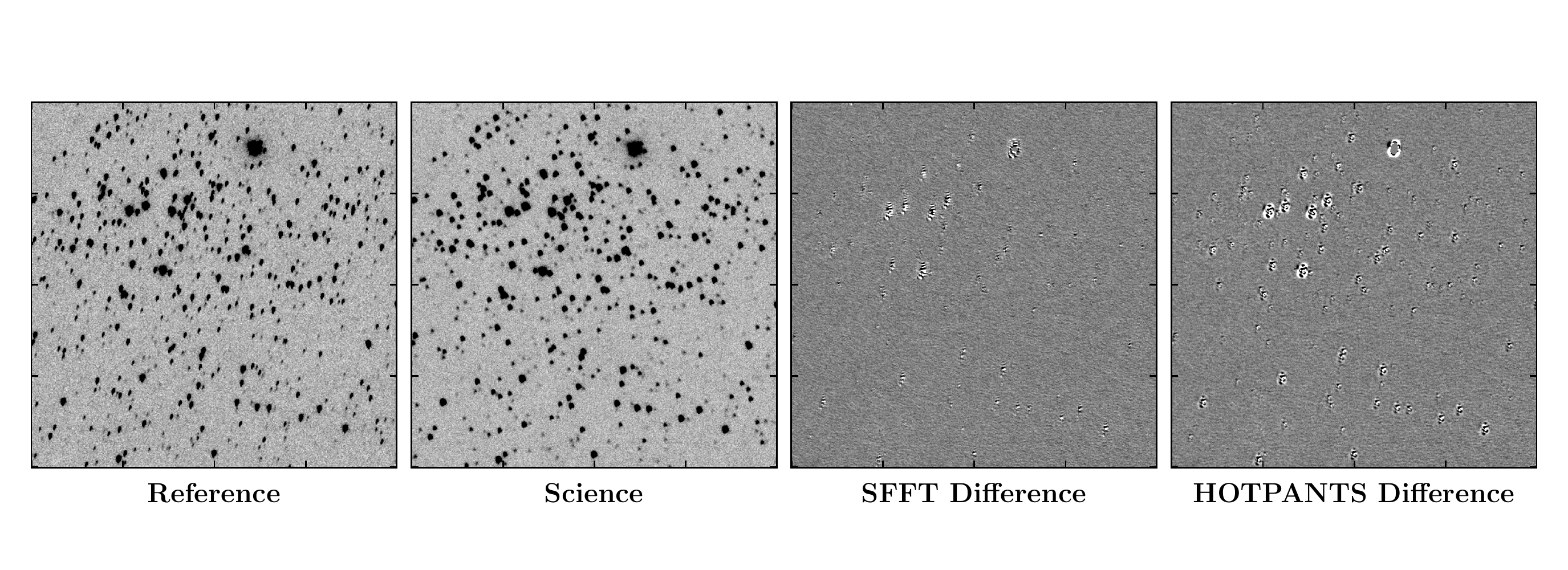}
    \caption{\label{fig:TMTS_example} {Image subtraction performance test using the TMTS images. The figure, from left to right, shows the input reference image $\textit{TMTS-REF}$, science image $\textit{TMTS-SCI}$ and resulting difference images by SFFT and {\tt\string HOTPANTS}.}}
\end{figure*}

It is also essential to investigate the performance of SFFT subtraction on the ordinary sparse sky fields that are more common in extragalactic transient surveys.
For comparison, we additionally employed the widely-used image subtraction software {\tt\string HOTPANTS} as a baseline method.
As one of specific implementations of \citet{AL98}, {\tt\string HOTPANTS} adopts Gaussian basis functions to construct the convolution kernel. 
In real space, the use of DBFs has been shown to yield better performance than Gaussian basis functions in terms of photometric accuracy \citep{Albrow09}.
Here we present the comparisons to {\tt\string HOTPANTS} to confirm that the SFFT, though working in Fourier space, will still inherit the advantage of DBFs over Gaussian basis functions.
The parameters of {\tt\string HOTPANTS} used for this section are listed in Table~\ref{tab:hp_param}.
Recall that $\textit{sparse-flavor}$ SFFT restricts the calculations of convolution kernels on a pre-selected set of sources, this source list used by SFFT will be shared by {\tt\string HOTPANTS} for the sake of fairness. 
Besides, both methods always apply the consistent size of convolution kernels and use the same polynomial degree of spatial variation for image subtraction.
We selected 25 DECam observations with the same pointing but diverse seeing conditions from our DECamERON survey \citep{DECamERON_Mould17} (see Table~\ref{tab:test_data_info}).
With the set of DECam images, we conduct two types of tests to check the subtraction cleanness and photometric accuracy, respectively. 

For the subtraction cleanness test, we use the image $\textit{DECam-SREF}$ as the shared reference, having the best seeing among all of them. 
The remaining 23 images aliased by initial $\textit{DECam-OBS}$ are regarded as science images in this test.
Image subtractions are performed for the sequence of science images with a shared reference using $\textit{sparse-flavor}$ SFFT and {\tt\string HOTPANTS}, respectively. An example of the subtraction results is shown in Figure~\ref{fig:DECam_example} from the observation $\textit{DECam-OBS02f}$. 
For both methods, the calculations involved in parameter-solving have been restricted to the same pre-selected sources, and the minimization aims at optimal subtraction on these sources as clean as possible.
It is useful to give a close-up view of subtraction performance specifically for the selection. For each image in Figure~\ref{fig:DECam_example}, we cut a small thumbnail image around each selected source and combine the cutouts into a grid to be a new synthetic image, as shown in Figure~\ref{fig:DECam_sample}. Note that the source selection, in this case, is more than 1100, so we only exhibit a subset of size 64 from the bright end of the complete set for a clear display.
One can notice that the most conspicuous subtraction-induced artifacts have a dipole-like pattern which is especially prominent for the bright sources. The pattern can be found in both difference images, but much less profound in SFFT subtractions than in {\tt\string HOTPANTS} subtractions.
It is an expected improvement as the SFFT kernel solution with DBFs allows more degrees of freedom than {\tt\string HOTPANTS} using Gaussian basis functions.

To further verify such an improvement is real in a statistical sense, here we defined a naive quantifiable metric to describe the subtraction cleanness and show the performance comparisons with the test data.
Note that the subtraction-induced artifacts on difference images are usually detectable by {\tt\string SEXTRACTOR} like other real transient and variable objects. Generally, a huge number of artifacts in transient surveys will survive until an AI-based stamp classifier recognizes them. 
This fact makes it possible to use a simple dichotomous metric to describe the subtraction cleanness. A given source is either $\textit{clean-subtracted}$ or not according to the existence of {\tt\string SEXTRACTOR} detection of its subtraction residuals.
Now we can investigate how the probability of $\textit{clean-subtraction}$ for a given set of sources is related to the employed image subtraction method.
For transient surveys, it is essential to probe the subtraction performance over the galaxies in the field, which are potential hosts of transient events. When a transient emerges at a position very close to its host galaxy core, the subtraction-induced artifacts from the galaxy can severely hinder the discovery and photometric measurement of the transient.

Our deep DECam observations in the test (3 square degrees, down to ~23.5 mag in i-band) have covered a vast number of galaxies with various morphological types. It is interesting to make statistics for the probability of $\textit{clean-subtraction}$ for those sources identified as extended galaxies (i.e., Tractor morphological types DEV, EXP, and REX) in the Legacy Survey \citep{LegacySurvey}.
The check radius from the galaxy core for searching subtraction-induced artifacts using {\tt\string SEXTRACTOR} is specified as 11 pixels ($\sim$ 2 times median FWHM for DECam).
Figure~\ref{fig:GalSub} shows the rate of $\textit{clean-subtraction}$ as a function of the magnitude of the examined galaxy. As expected, the brighter galaxies have a higher probability of leaving detectable subtraction-induced artifacts. However, the SFFT method is more likely to achieve $\textit{clean-subtraction}$ than {\tt\string HOTPANTS}.

For the photometric accuracy test, we use the image $\textit{DECam-PTAR}$ as the PSF target, which has a median level seeing among all selected DECam observations. The remaining 23 images aliased by initial $\textit{DECam-OBS}$ are convolved to achieve homogeneous PSF using $\textit{sparse-flavor}$ SFFT and {\tt\string HOTPANTS}, respectively. 
We derive the light curves of the field stars by conducting photometry on the sequence of convolved images with a fixed aperture. The photometric accuracy measured by the light curves can work as a metric to assess the PSF matching quality. We investigate the aperture photometry of the point sources classified as Tractor morphological type PSF. Figure~\ref{fig:PhotAcc} shows the RMS error of the derived light curves as a function of the star brightness. 
As the photometric uncertainty from SFFT is systematically lower than that from {\tt\string HOTPANTS} especially at the bright end, the improvement in PSF matching is again confirmed.

Figure~\ref{fig:TMTS_example} shows an example with TMTS data. The subtraction artifacts generated by SFFT and {\tt\string HOTPANTS} are not identical. {\tt\string HOTPANTS} produces relatively circular and symmetric patterns, while SFFT shows more random residuals. This obviously originates from the different base kernel construction in SFFT and {\tt\string HOTPANTS}. SFFT uses a $\delta$-basis kernel which is more flexible, while {\tt\string HOTPANTS} utilizes Gaussian functions which possess a higher intrinsic symmetry about the kernel center.

\section{Computational Performance of SFFT} \label{sec:sfft_comput_perf}

\begin{figure*}[ht!]
    \gridline{\fig{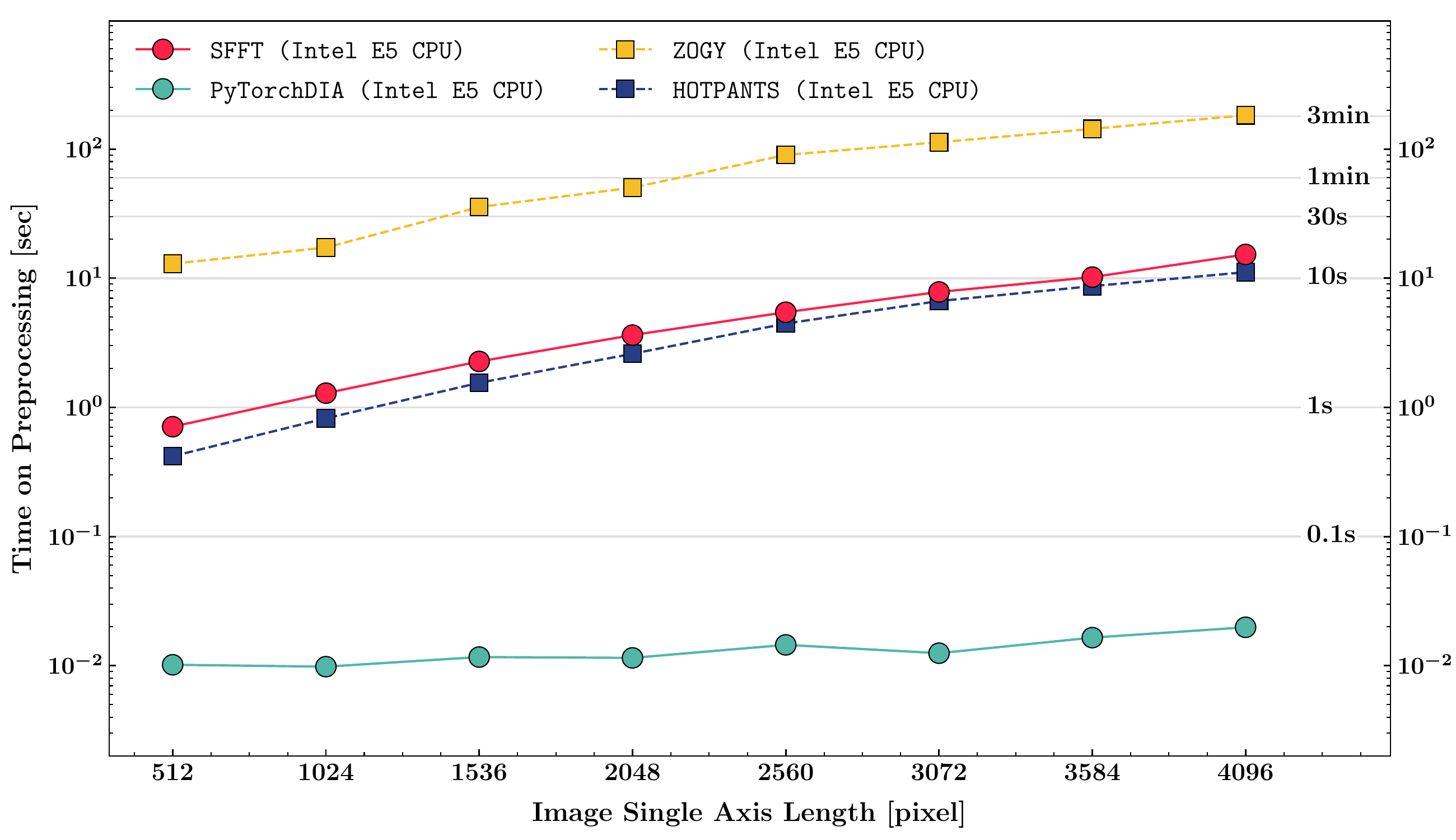}{0.49\textwidth}{(a) server $\mathbf{I}$}
              \fig{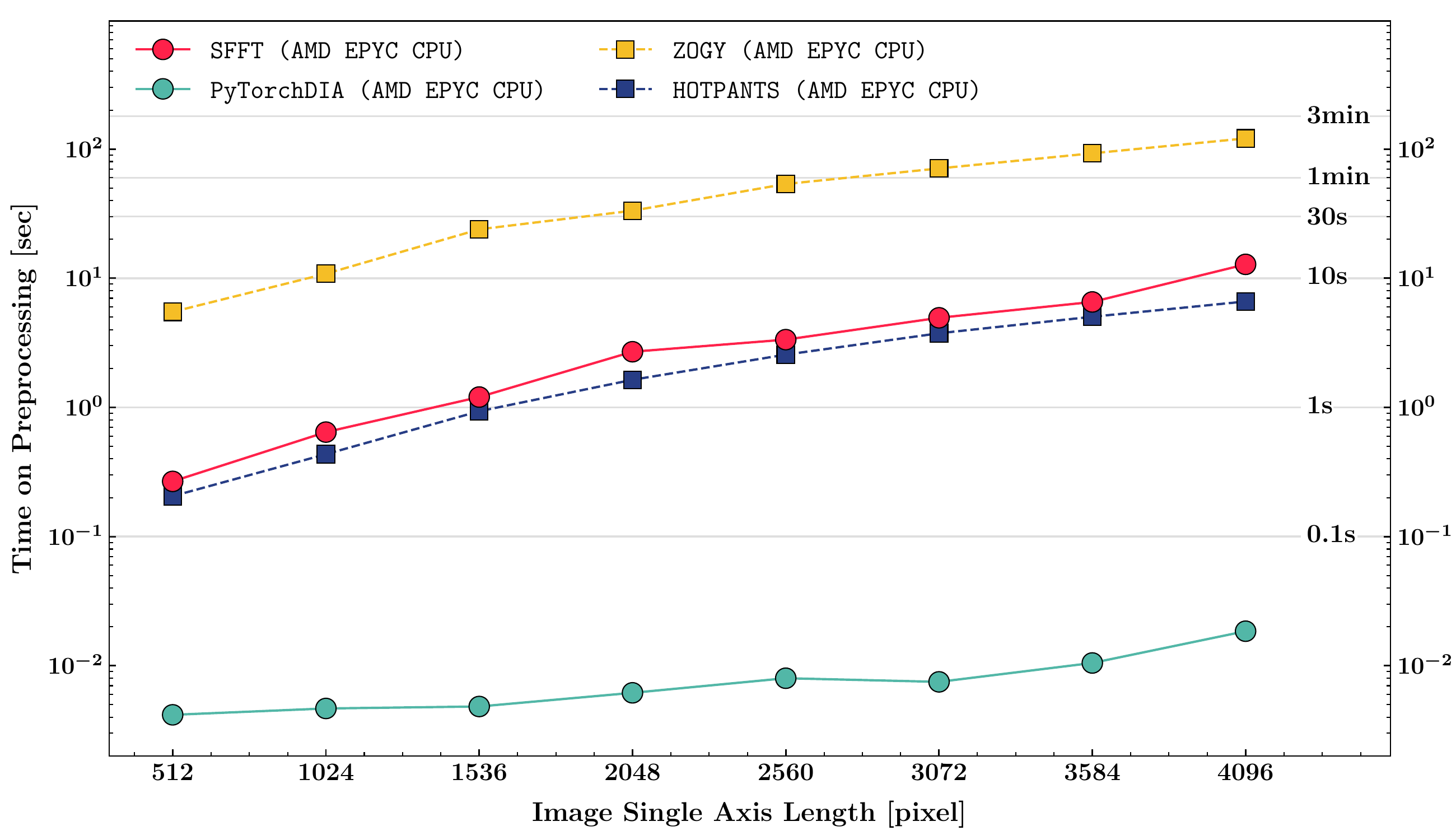}{0.49\textwidth}{(b) server $\mathbf{II}$}}
    \caption{\label{fig:PreSubtractTime_Comparison} Computing time on preprocessing versus the image size of input data for different methods. The curves with different colors represent the measured time spent on pre-subtraction procedures using different methods on server $\mathbf{I}$ (left) and server $\mathbf{II}$ (right). Note that each test using SFFT, {\tt\string HOTPANTS} or {\tt\string PyTorchDIA} has been repeated with a specified kernel size in each run. Given that the kernel size is irrelevant to preprocessing, we simply use the average time in the figure.}
\end{figure*}

\begin{figure*}[ht!]
    \gridline{\fig{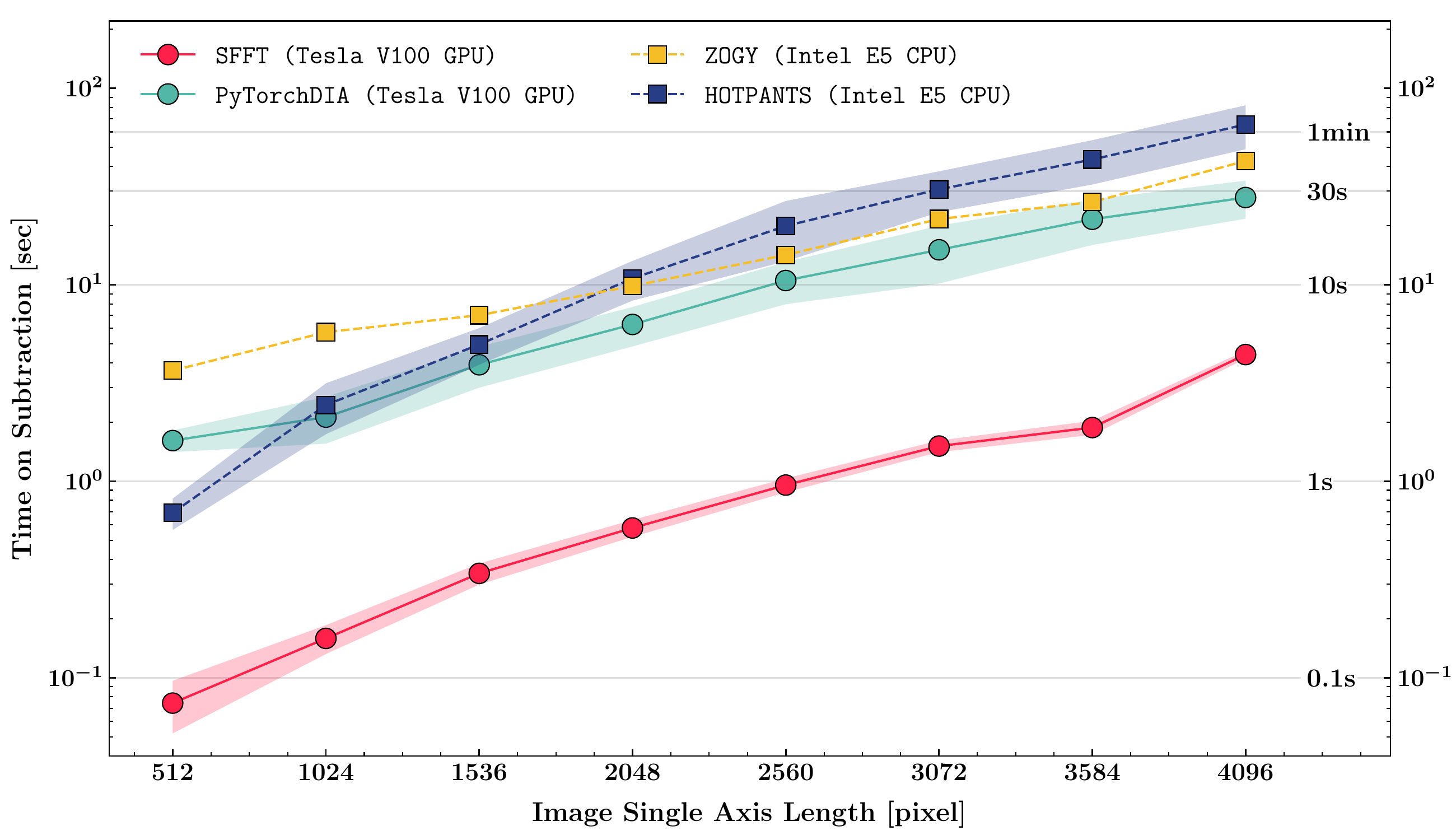}{0.49\textwidth}{(a) server $\mathbf{I}$}
              \fig{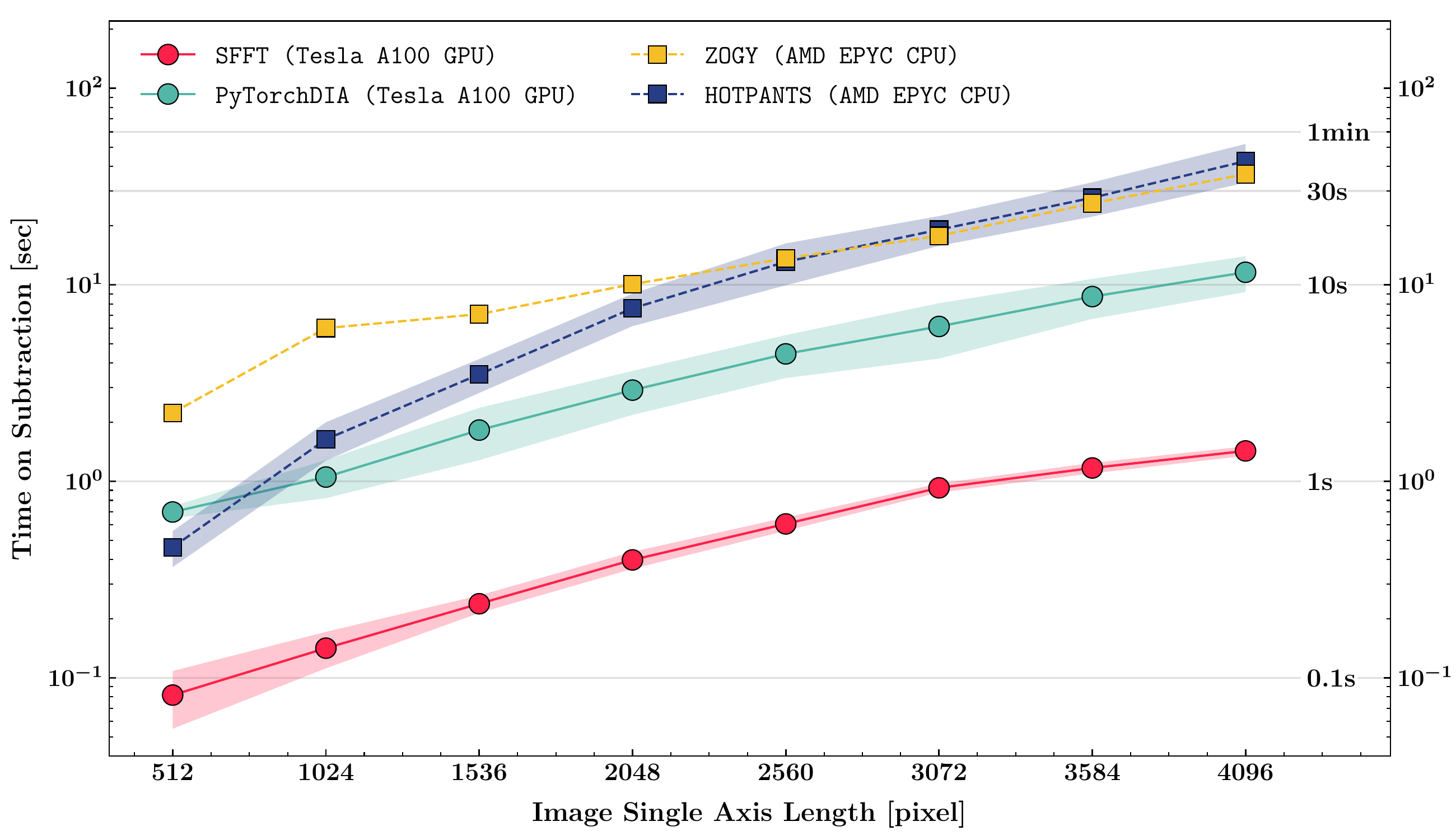}{0.49\textwidth}{(b) server $\mathbf{II}$}}
    \caption{\label{fig:SubtractTime_Comparison} Computing time on image subtraction versus the image size of input data for different methods. The curves with different colors represent the measured time spent on subtraction using different methods on server $\mathbf{I}$ (left) and server $\mathbf{II}$ (right). For SFFT, {\tt\string HOTPANTS} or {\tt\string PyTorchDIA}, the data points in the curves represent the average time measured from the multiple runs, and the shaded areas indicate the corresponding 1$\sigma$ standard deviation.}
\end{figure*}

The most remarkable feature of SFFT is that the image subtraction in Fourier space can be efficiently parallelized in GPU devices. With the simplification described in Appendix~\ref{App:simplify}, we have reduced the computation-intensive kernel determination as discrete Fourier transforms (DFTs) and element-wise matrix operations, which are exceptionally suitable for GPU graphic multi-processors.
This section will show the excellent computing performance of the SFFT algorithm in comparison with several existing image subtraction methods, including {\tt\string HOTPANTS}, {\tt\string ZOGY}\footnote{\url{https://github.com/pmvreeswijk/ZOGY}}, and {\tt\string PyTorchDIA}\footnote{\url{https://github.com/jah1994/PyTorchDIA}}.
Note that the comparisons only focus on the limited specific implementations of image subtraction algorithms. One should not overinterpret the computing costs given in this section to represent the best possible computational performances of the general algorithms.
Unless specified otherwise, the parameters of those softwares used for this section are listed in Tables~{\ref{tab:sfft_param},\ref{tab:hp_param},\ref{tab:zogy_param},\ref{tab:ptdia_param}}.

\begin{table*}[ht!]
    \begin{center}
    \tablenum{2}
    \begin{tabular}{cccccccccccc}
        \Xcline{1-8}{1.3pt}
        Method & \multicolumn{2}{c}{Time Cost on Preprocessing (sec)} & & \multicolumn{4}{c}{Time Cost on Subtraction (sec)} \\
        \cline{2-3}
        \cline{5-8}
        & Intel E5 & AMD EPYC & & Intel E5 & AMD EPY & Tesla V100 & Tesla A100 \\
        & CPU & CPU & & CPU & CPU & GPU & GPU \\
        \hline
        SFFT & 5.75 $\pm$ 0.26  & 3.35 $\pm$ 0.15 & & - & - & 1.61 $\pm$ 0.10 & 0.97 $\pm$ 0.06 \\
        PyTorchDIA & 0.11 $\pm$ 0.01 & 0.16 $\pm$ 0.07 & & - & - & 15.87 $\pm$ 3.25 & 6.45 $\pm$ 1.36 \\
        HOTPANTS & 4.37 $\pm$ 0.85 & 2.43 $\pm$ 0.39 & & 26.39 $\pm$ 8.50 & 15.39 $\pm$ 3.49 & - & - \\
        ZOGY & 63.59 $\pm$ 6.42 & 42.28 $\pm$ 3.74 & & 62.83 $\pm$ 9.16 & 41.01 $\pm$ 5.25 & - & - \\
        \Xcline{1-8}{1.3pt}
        \end{tabular}
        \caption{\label{tab:speed_decam} Comparison of the computing time on preprocessing and subtraction measured from the tests on DECam data using different methods.}
    \end{center}
\end{table*}

It is useful to demonstrate the computing performance of the image subtraction methods by testing them with a variety of data on different computation platforms. 
We select a set of testing input data with diverse image sizes, seeing conditions, and sky areas.
The tests are performed on two computing servers with different hardware configurations.
One server (denoted by $\mathbf{I}$) is equipped with Intel(R) Xeon(R) CPU E5-2630 (2.20 GHz) and one NVIDIA Tesla V100 GPU, while the other server (denoted by $\mathbf{II}$) has more advanced configurations with AMD EPYC CPU 7542 and one NVIDIA Tesla A100 GPU. 
Note that the time spent on preprocessing steps can also be considerable. We additionally record the computing cost of the pre-subtraction procedures in our tests. In this section, we conduct two types of tests using TMTS data and DECam data, respectively.

To explore the relationship between the computing time and input image size, we extract sub-images of different sizes from the original TMTS image-pair (i.e., $\textit{TMTS-REF}$ and $\textit{TMTS-SCI}$).
We then perform image subtractions for the trimmed image-pairs on the two servers using the four tested methods. For SFFT, {\tt\string HOTPANTS} and {\tt\string PyTorchDIA}, in order to test if the computational cost is sensitive to the size of the convolutional kernel, we run the image subtractions repeatedly with a specified kernel size from 15 to 25 pixels in each run (six runs in total).

Figure~\ref{fig:PreSubtractTime_Comparison} shows the computing time on the preprocessing as a function of input image size. In general, larger input images consume more computing time on preprocessing. 
Among the tested methods, {\tt\string ZOGY} requires the longest preprocessing time. 
For {\tt\string ZOGY}, the preprocessing is composed mainly of fitting the PSF models for input images and calculating their flux ratio as well as the astrometric registration noises.
The preprocessing time of SFFT and {\tt\string HOTPANTS} are comparable and much shorter than {\tt\string ZOGY}. For these two methods, the goal of the pre-subtraction procedures is to select an optimal set of sources for kernel determination. 
Like in Section~\ref{subsec:sparse-tests}, we have forced {\tt\string HOTPANTS} to use the pre-selected set given by SFFT following Section~\ref{subsec:preproc-sparse}. Note that SFFT needs further to perform an image-masking based on the pre-selected set while {\tt\string HOTPANTS} does not, that is why the preprocessing time of SFFT is slightly longer than {\tt\string HOTPANTS}.
For {\tt\string PyTorchDIA}, the preprocessing only contains reading the input FITS images. Thus, it can achieve minimal preprocessing time. We note that the robust loss function used in {\tt\string PyTorchDIA} is designed to be insensitive to the $\textit{distraction-sources}$, as a result, no source selection or image-masking is involved in {\tt\string PyTorchDIA}.

Figure~\ref{fig:SubtractTime_Comparison} presents the computing time on the image subtraction versus the input image size. Unsurprisingly, larger images generally require more computing time for subtraction. 
The computing speeds for {\tt\string HOTPANTS} and {\tt\string ZOGY} are broadly comparable, with a typical cost of 40-90s for 4K images. 
By contrast, the GPU-powered method {\tt\string PyTorchDIA} brings considerable speed-up. For 4K images, the computing time is around 30s and 10s on the server $\mathbf{I}$ and $\mathbf{II}$, respectively. 
Note that {\tt\string PyTorchDIA} only solves a spatially invariant kernel while other tested methods can accommodate spatially varying PSF.
Remarkably, SFFT is almost an order of magnitude faster than {\tt\string PyTorchDIA}. For 4K images, the computing time is close to 4s and 1s on the server $\mathbf{I}$ and ($\mathbf{II}$), respectively. As shown in Figure~\ref{fig:SubtractTime_Comparison}, the subtraction speed of SFFT is less affected by the kernel size compared to {\tt\string HOTPANTS} and {\tt\string PyTorchDIA}.

The second type of test aims to explore the computing expense on a set of images with diverse seeing conditions covering different sky areas. 
We randomly select 100 CCD images from the 23 DECam observations aliased by initial $\textit{DECam-OBS}$ (see Table~\ref{tab:test_data_info}). Note that one DECam exposure produces 62 CCD images, each with a size of 2046$\times$4094 pixels. 
As shown in Table~\ref{tab:test_data_info}, the DECam data were obtained under varying seeing conditions. Although the DECam exposures listed in Table~\ref{tab:test_data_info} have similar telescope pointing, the arbitrarily selected CCD images can, to some extent, cover a variety of distributions of the observed stellar objects. 
In this test, the selected 100 images are regarded as science images, and we use the corresponding CCD images of $\textit{DECam-SREF}$ as their reference images. We perform image subtractions for the 100 image-pairs on the two servers using the four tested methods. Unlike the above test, we do not specify a given kernel size but leave it automatically determined according to the seeing condition. 
Table~\ref{tab:speed_decam} summarizes the time costs on preprocessing and image subtraction for the tested DECam dataset. The results are broadly consistent with the first test. Again, SFFT shows a distinct advantage in the subtraction speed over other tested methods. Given image size, SFFT is the most stable method on subtraction time, with a percentage variation $\sim6\%$ (the typical value is $20\%$ for other methods).

We note that the software {\tt\string HOTPANTS} also allows users to perform image subtraction without a pre-selected source list. In such an automatic mode, {\tt\string HOTPANTS} will accomplish the source selection using its built-in functions. 
One may wonder if the automatic mode can be faster than the manual mode used in this paper. Hence, we additionally perform the image subtraction on the 100 image-pairs using the automatic mode {\tt\string HOTPANTS}. It turns out that the total time (preprocessing plus subtraction) for the two modes is very close, with the automatic mode having a slightly higher variation.

For the crowded fields, the computational cost of SFFT will not rise to a level higher than the cost shown in this section. For preprocessing, $\textit{crowded-flavor}$ SFFT involves much simpler operations than $\textit{spase-flavor}$ (see Section~\ref{subsec:preproc-crowded}). On the other hand, signal-dominated input images will not prolong the computing time on image subtraction in Fourier space.

\vspace{0.1cm}
\begin{table}[ht!]
    \begin{center}
    \tablenum{3}
    \begin{tabular}{ccccccccccc}
        \Xcline{1-5}{1.3pt}
        HOTPANTS & \multicolumn{2}{c}{Total Time Cost (sec)} \\
        \cline{2-3}
        Mode & Intel E5 & AMD EPYC \\
        \hline
        Manual (this work) & 30.76 $\pm$ 8.81 & 17.82 $\pm$ 3.62 & & \\
        Automatic & 30.75 $\pm$ 10.54 & 17.71 $\pm$ 4.40 & &\\
        \Xcline{1-5}{1.3pt}
        \end{tabular}
    \caption{\label{tab:speed_autohp} Comparison of the total computing time measured from the tests on DECam data using {\tt\string HOTPANTS} in two modes.}
    \end{center}
\end{table}

\section{Limitations and Future Works} \label{sec:sfft_limfuture}

\begin{table*}
    \begin{center}
    \tablenum{4}
    \begin{tabular}{ccccccccc}
        \Xcline{1-8}{1.3pt}
        Method & \multicolumn{3}{c}{Convolution Kernels} & & \multicolumn{3}{c}{Accommodate Spatial Variations} \\
        \cline{2-4}
        \cline{6-8}
        & Gaussian-function & $\delta$-function & Regularized & & Convolution & Photometric  & Differential \\
        & basis & basis & Kernel &  & Kernels & Scale Factor & Background \\
        \hline
        SFFT & \xmark & \cmark & \xmark & & \cmark & \cmark & \cmark   \\
        PyTorchDIA  & \xmark & \cmark & \xmark & & \xmark & \xmark & \cmark  \\
        HOTPANTS & \cmark & \xmark & \xmark & & \cmark  & \xmark & \cmark    \\
        ZOGY & - & - & - & &  \cmark & \cmark & - \\
        \Xcline{1-8}{1.3pt}
        Method & \multicolumn{7}{c}{Kernel Determination} \\
        \cline{2-8}
        & Require PSF & Require Isolated & Rely on Rectangle & & Weighting by & Construct Least & Symmetric to \\
        & Knowledge & Objects &  Sub-areas &  &  Pixel Noise & -squares Matrix & Image Exchange \\
        \hline
        SFFT & \xmark & \xmark & \xmark & & \xmark & \cmark & \xmark   \\
        PyTorchDIA  & \xmark & \xmark & \xmark & & \cmark & \xmark & \xmark  \\
        HOTPANTS & \xmark & \cmark & \cmark & & \xmark & \cmark & \xmark    \\
        ZOGY & \cmark & \cmark & \cmark & & - & \xmark & \cmark \\
        \Xcline{1-8}{1.3pt}
        Method & \multicolumn{3}{c}{Preprocessing before Subtraction} & & \multicolumn{3}{c}{Computing Performance of Subtraction} \\
        \cline{2-4}
        \cline{6-8}
        & Stamp Selection & PSF & Calculate &  & Implemented & Built-in CPU & Speed Sensitive \\
        & /Image Masking & Modeling & Flux Ratio &  & for GPU & Multithreading & to Kernel Size \\
        \hline
        SFFT & \cmark & \xmark & \xmark & & \cmark & - & \xmark   \\
        PyTorchDIA  & \xmark & \xmark & \xmark & & \cmark & - & \cmark  \\
        HOTPANTS & \cmark & \xmark & \xmark & & \xmark  & \xmark & \cmark    \\
        ZOGY & - & \cmark & \cmark & &  \xmark & \cmark & - \\
        \Xcline{1-8}{1.3pt}
        \end{tabular}
        \tablecomments{For {\tt\string PyTorchDIA}, the kernel regularization has been mentioned in \citet{PyTorchDIA21} whereas no regularization has been implemented in the software {\tt\string PyTorchDIA}; For {\tt\string HOTPANTS}, the kernel determination relies on a set of stamps encompassing isolated objects. Note that requiring isolated objects is not a feature of \citet{AL98}. Moreover, pixel noise is not used in the matrix calculation in {\tt\string HOTPANTS} though formulated in \citet{AL98}; For {\tt\string ZOGY}, all the convolutions involved in image subtraction are determined by pre-calculated PSFs. Generally, constructions of the PSF models are based on a sample of rectangle sub-images (e.g., {\tt \string SExtractor} VIGNETS) centered at isolated bright point sources.}
    \caption{\label{tab:method_comp} Characteristics of SFFT compared with other existing image subtraction methods.}
    \end{center}
\end{table*}

To get a clear picture of the current SFFT and find its limitations, we summarize a variety of features from different perspectives for SFFT and other existing image subtraction methods, including {\tt\string PyTorchDIA}, {\tt\string HOTPANTS} and {\tt\string ZOGY} in Table~\ref{tab:method_comp}.
Although a general algorithm \citep[e.g.,][]{AL98} can have multiple existing implementations, the scope of the comparisons in Table~\ref{tab:method_comp} is restricted: we have only considered a specific implementation for each algorithm. 
Some characteristics of a specific implementation may not necessarily be the intrinsic features of its underlying algorithm (see the note attached to Table~\ref{tab:method_comp}).

Although SFFT uses the state-of-the-art DBFs as kernel basis functions, the current version does not contain an effective mechanism to aginst the overfitting problem of DBFs. In future work, we may try to incorporate existing solutions into SFFT, e.g., \citet{Becker12,Bramich16,iPTF_Masci17}. Moreover, unlike \citet{AL98,Bramich08,PyTorchDIA21}, the kernel determination in SFFT is not weighted by pixel noise, which may introduce a possible bias towards the brighter pixels. It looks tricky to design a weighting scheme in Fourier space. 
Note that some practical implementations, including {\tt\string HOTPANTS} and \citet{Miller08}, also ignore the pixel weights in kernel determination.

As discussed in \citet{ZOGY}, the approach proposed by \citet{AL98} is not symmetric, i.e., changing the convolution direction of image subtraction will generally result in inconsistent difference images. Consequently, SFFT has some limitations inherited from the nature of this kind. 
In particular, when the direction matches an image to another one with better seeing, the deconvolution effect makes its way into the subtraction to amplify the noise and strengthen noise correlation on the difference image. In many cases, one can steer clear of the deconvolution effect by simply adjusting the direction. However, this strategy is not a panacea; e.g., it may be insufficient when the observations have elliptic-like PSF with varying orientations \citep{ZOGY}.

Another limitation of the current SFFT is related to the computing cost of preprocessing. The image-masking scheme of SFFT introduced in Section~\ref{sec:sfft_implement} provides a safe and generic approach that has been proved to be reliable through extensive tests by multiple transient surveys. However, its computing cost might become a potential encumbrance for those projects that pursues extremely rapid transient detection. 
Fortunately, there is ample space for speed-up in a given survey program. One can modify the SFFT code to design a particular image-masking function for optimizing its overall computing expense. 
For example, the {\tt\string SEXTRACTOR} catalogs used for the source selection can be provided by preceding modules (e.g., astrometric calibration) in the pipeline.
When processing time-series data taken from the same pointing, one can skip the repeated photometry for the shared reference image to save the preprocessing time.
One may also consider using a single pre-defined mask (fixed or seeing-related) for all images in the time series.
We note that {\tt\string PyTorchDIA} can take very little time on preprocessing as it designs a specific loss function insensitive to outlying values, which provides an alternative to suppress the impact from $\textit{distraction-sources}$ amid the kernel determination.

In the current implementations of SFFT, the subtraction has two flavors that make image-masking in different ways. However, there is no built-in discriminator for sparse fields and crowded fields in SFFT. One may specify the adopted flavor for each field to be processed. In future work, we may consider developing a more unified way to conduct image-masking rather than dividing the observations into two subgroups. 
Besides, the limited memory amount of GPU devices can constrain the applicability of the GPU-based SFFT to very large images. To allows for large image cases, we have provided a Numpy-based backend for SFFT without GPU requirement. This is also helpful for users who do not have available GPU resources.

\section{Summary and Conclusions}

SFFT is a novel image subtraction algorithm formulated in Fourier space.
Like in the classic framework of \citet{AL98}, SFFT addresses the PSF discrepancy between two images by the convolution with pixelized kernels that are decomposed into pre-defined basis functions. 
In our work, we adopted the $\delta$-function basis proposed by \citet{Miller08} for kernel flexibility and minimal hyper-parameters.
SFFT solves the least-squares problems of image subtraction in the Fourier domain, and it allows for space-varying PSF, photometric scaling, and sky background modeling across the entire field-of-view. 
In particular, SFFT can reduce the computation-intensive kernel determination to FFTs and element-wise matrix operations. By leveraging CUDA-enabled GPU acceleration, SFFT empowers a great leap forward in computational performance of astronomical image subtraction with around one order of magnitude speed gain.

We demonstrated in this paper the power of the method on real astronomical images taken by a variety of telescopes. We show that SFFT can accommodate imaging data with diverse characteristics, including sparse fields, crowded fields, and under-sampled PSFs. 
The examples of SFFT for image subtraction are presented in Section~\ref{sec:sfft_examples}. 
We confirm that SFFT yields a high quality of PSF-matching when compared with the widely-used image subtraction software {\tt\string HOTPANTS} using Gaussian kernel basis for models of the PSFs. 
A comprehensive investigation of the computing performance of SFFT is presented in Section~\ref{sec:sfft_comput_perf}. 
We find that SFFT is not only faster compared to other existing image subtraction methods but is also more stable and efficient.

We also explored the limitations of the current SFFT implementation in Section~\ref{sec:sfft_limfuture}. 
One prominent weakness of the current SFFT is that the kernel solutions may suffer from the overfitting problem due to the high degree of freedom of DBFs.
A few approaches have been proposed to alleviate the overﬁtting issue of DBFs, e.g., regularizing the kernel solutions \citep{Becker12}. However, the DBFs employed by the current SFFT are unregularized. The regularization techniques will be included in future works. In the meantime, proper image masking is required in the current version of SFFT to identify the pixels that are not correctly modeled by SFFT. 
In principle, SFFT also provides a generic and robust function to perform preprocessing of the data, which has been extensively tested with data from various transient surveys. 
In contrast to the high speed of the image subtraction, however, the computing performance of the generic preprocessing is less remarkable. 
Given a particular time-domain program, we believe there is plenty of room for further optimization of the computing expense on the preprocessing.

We conclude that SFFT has the potential to be the optimal image subtraction engine for future time-domain surveys with massive data flows that require differential image subtractions for transient detection and precision differential photometry.

\acknowledgments
L.H. acknowledges the supports from the National Natural Science Foundation of China (11761141001) and the Major Science and Technology Project of Qinghai Province (2019-ZJ-A10). L.W., X.C. and J.Y. acknowledge supports from an NSF grant AST1817099.
We thank Bo Yu and Tianrui Sun for useful discussions on mathematical derivation and package developments.
The authors are also grateful to an anonymous referee whose report has significantly improved this manuscript.

\software{
    Astropy \citep{astropy13}, 
    SciPy \citep{2020SciPy-NMeth}, 
    Numpy \citep{2020NumPy-Array},
    MatPlotLib \citep{Matplotlib},
    CuPy \citep{CuPy},
    PyCuda \citep{PyCuda},
    Scikit-CUDA \citep{Scikit-CUDA},
    Numba \citep{Numba},
    SExtractor \citep{SExtractor},
    SWarp \citep{SWarp},
    PSFEx \citep{PSFEx},
    HOTPANTS \citep{HOTPANTS15},
    ZOGY \citep{ZOGY},
    PyTorchDIA \citep{PyTorchDIA21},
    SFFT \citep{sfft_zenodo}
}

\clearpage
\appendix

\section{The Approximation in SFFT} \label{App:approx}

Here we will prove the approximation applied to Equation~(\ref{eqn:sfft_eq10}). Given any image R and a kernel Q, where Q has a much smaller size than R as it is in most astronomical applications.

\begin{equation}
\label{eqn:A1}
T^{\rho\epsilon}(R \circledast Q) \approx R^{\rho\epsilon} \circledast Q.
\end{equation}

Writing the Left Hand Side ($\textbf{LHS}$) and Right Hand Side ($\textbf{RHS}$) as follows,

\begin{equation}
\label{eqn:A2}
\begin{split}
\textbf{LHS}(r, s) &= r^\rho s^\epsilon \sum_{c = - w_0} ^{w^\prime_0}  \sum_{d = - w_1} ^{w^\prime_1} Q(c, d) R_\mathfrak{p}(r-c, s-d)\\
&= \sum_{c = - w_0} ^{w^\prime_0}  \sum_{d = - w_1} ^{w^\prime_1} Q(c, d) [r^\rho s^\epsilon R_\mathfrak{p}(r-c, s-d)].
\end{split}
\end{equation}

\begin{equation}
\label{eqn:A3}
\begin{split}
\textbf{RHS}(r, s) &= \sum_{c = - w_0} ^{w^\prime_0}  \sum_{d = - w_1} ^{w^\prime_1} Q(c, d) R_\mathfrak{p}^{\rho\epsilon}(r-c, s-d)\\
&= \sum_{c = - w_0} ^{w^\prime_0}  \sum_{d = - w_1} ^{w^\prime_1} Q(c, d)  (T^{\rho\epsilon} R)_\mathfrak{p} (r-c, s-d).
\end{split}
\end{equation}

Note that the indices of kernel $Q$ are modified with origin at its center pixel, and the subscript $\mathfrak{p}$ is employed to indicate the periodic extension.

For $r$ and $s$, $(r-c, s-d)$ is a vector within the image frame and $(r-c, s-d) \approx (r, s)$ for nearly all possible $c$ and $d$.

\begin{equation}
\label{eqn:A4}
\begin{split}
\textbf{LHS}(r, s) &= \sum_{c = - w_0} ^{w^\prime_0}  \sum_{d = - w_1} ^{w^\prime_1} Q(c, d) [r^{\rho} s^{\epsilon} R(r-c, s-d)]
\\
&\approx \sum_{c = - w_0} ^{w^\prime_0}  \sum_{d = - w_1} ^{w^\prime_1} Q(c, d) [(r-c)^{\rho} (s-d)^{\epsilon} 
\\
&\;\;\;\;\;\;R(r-c, s-d)] = \textbf{RHS}(r, s).
\end{split}
\end{equation}

For the particular case $\rho=\epsilon=0$, it becomes a rigorous equation rather than an approximation.

\section{Calculation Simplification} \label{App:simplify}

The $\delta$-function basis facilitates further simplifications.
Equation~(\ref{eqn:sfft_eq19}) will be re-expressed as 

\begin{equation}
\label{eqn:B5}
\begin{split}
A_{\bar{i}\bar{j}\bar{\alpha}\bar{\beta} ij\alpha\beta} &= -\Omega_{\bar{i}\bar{j} ij}(\bar{\alpha}, \bar{\beta}) - \Omega_{\bar{i}\bar{j} ij} (-\alpha, -\beta) 
\\
&\;\;\;\; + \Omega_{\bar{i}\bar{j} ij} (\bar{\alpha} - \alpha, \bar{\beta} - \beta) + \Omega_{\bar{i}\bar{j} ij} (0, 0)
\\
&\;\;\;\;\;\;\;\;\;\;\textbf{\textit{if }} (\bar{\alpha}, \bar{\beta}) \neq \vec{0} \textbf{\textit{ and }} (\alpha, \beta) \neq \vec{0}
\\
A_{\bar{i}\bar{j}\bar{\alpha}\bar{\beta} ij\alpha\beta} &= \Omega_{\bar{i}\bar{j} ij} (-\alpha, -\beta) - \Omega_{\bar{i}\bar{j} ij} (0, 0)
\\
&\;\;\;\;\;\;\;\;\;\;\textbf{\textit{if }} (\bar{\alpha}, \bar{\beta}) = \vec{0} \textbf{\textit{ and }} (\alpha, \beta) \neq \vec{0}
\\
A_{\bar{i}\bar{j}\bar{\alpha}\bar{\beta} ij\alpha\beta} &= \Omega_{\bar{i}\bar{j} ij} (\bar{\alpha}, \bar{\beta}) - \Omega_{\bar{i}\bar{j} ij} (0, 0)
\\
&\;\;\;\;\;\;\;\;\;\;\textbf{\textit{if }} (\bar{\alpha}, \bar{\beta}) \neq \vec{0} \textbf{\textit{ and }} (\alpha, \beta) = \vec{0}
\\
A_{\bar{i}\bar{j}\bar{\alpha}\bar{\beta} ij\alpha\beta} &= \Omega_{\bar{i}\bar{j} ij} (0, 0)
\\
&\;\;\;\;\;\;\;\;\;\;\textbf{\textit{if }} (\bar{\alpha}, \bar{\beta}) = \vec{0} \textbf{\textit{ and }} (\alpha, \beta) = \vec{0},
\end{split}
\end{equation}

where
\begin{equation}
\label{eqn:B6}
\begin{split}
\Omega_{\bar{i}\bar{j} ij} (\rho, \epsilon) &= \mathring{\Omega}_{\bar{i}\bar{j} ij} (\rho\bmod{N0}, \epsilon\bmod{N1})
\\
\mathring{\Omega}_{\bar{i}\bar{j} ij} &= \frac{1}{N_0N_1} \Re\;[\textbf{DFT}(\widehat{R^{\bar{i}\bar{j}}} (\widehat{R^{ij}})^*)].
\end{split}
\end{equation}

Equation~(\ref{eqn:sfft_eq20}) can be rewritten as 
\begin{equation}
\label{eqn:B7}
\begin{split}
B_{\bar{i}\bar{j}\bar{\alpha}\bar{\beta} pq} &= \Lambda_{\bar{i}\bar{j} pq}(\bar{\alpha}, \bar{\beta}) - \Lambda_{\bar{i}\bar{j} pq} (0, 0) \;\;\;\textbf{\textit{if}} (\bar{\alpha}, \bar{\beta}) \neq \vec{0}
\\
B_{\bar{i}\bar{j}\bar{\alpha}\bar{\beta} pq} &= \Lambda_{\bar{i}\bar{j} pq} (0, 0) \;\;\;\;\;\;\;\;\;\;\;\;\;\;\;\;\;\;\;\;\;\;\;\;\textbf{\textit{if}} (\bar{\alpha}, \bar{\beta}) = \vec{0},
\end{split}
\end{equation}
where
\begin{equation}
\label{eqn:B8}
\begin{split}
\Lambda_{\bar{i}\bar{j} pq} (\rho, \epsilon) &= \mathring{\Lambda}_{\bar{i}\bar{j} pq} (\rho\bmod{N_0}, \epsilon\bmod{N_1})
\\
\mathring{\Lambda}_{\bar{i}\bar{j} pq} &= \Re\;[\textbf{DFT}(\widehat{R^{\bar{i}\bar{j}}} (\widehat{T^{pq}})^*].
\end{split}
\end{equation}

Equation~(\ref{eqn:sfft_eq21}) is written as
\begin{equation}
\label{eqn:B9}
\begin{split}
\tilde{B}_{\bar{p}\bar{q} ij\alpha\beta} &= \Psi_{\bar{p}\bar{q} ij}(-\alpha, -\beta) - \Psi_{\bar{p}\bar{q} ij} (0, 0) \;\;\;\textbf{\textit{if}} (\alpha, \beta) \neq \vec{0}
\\
\tilde{B}_{\bar{p}\bar{q} ij\alpha\beta} &= \Psi_{\bar{p}\bar{q} ij} (0, 0) \;\;\;\;\;\;\;\;\;\;\;\;\;\;\;\;\;\;\;\;\;\;\;\;\;\;\;\;\;\;\textbf{\textit{if}}  (\alpha, \beta) = \vec{0},
\end{split}
\end{equation}
where
\begin{equation}
\label{eqn:B10}
\begin{split}
\Psi_{\bar{p}\bar{q} ij} (\rho, \epsilon) &= \mathring{\Psi}_{\bar{p}\bar{q} ij} (\rho\bmod{N_0}, \epsilon\bmod{N_1})
\\
\mathring{\Psi}_{\bar{p}\bar{q} ij} &= \Re\;[\textbf{DFT}(\widehat{T^{\bar{p}\bar{q}}}  (\widehat{R^{ij}})^*)].
\end{split}
\end{equation}
Equation~(\ref{eqn:sfft_eq22}) turns out to be
\begin{equation}
\label{eqn:B11}
\begin{split}
C_{\bar{p}\bar{q} pq} &= \Phi_{\bar{p}\bar{q} pq} (0, 0),
\end{split}
\end{equation}
where
\begin{equation}
\label{eqn:B12}
\begin{split}
\Phi_{\bar{p}\bar{q} pq} (\rho, \epsilon) &= \mathring{\Phi}_{\bar{p}\bar{q} pq} (\rho\bmod{N_0}, \epsilon\bmod{N_1})
\\
\mathring{\Phi}_{\bar{p}\bar{q} pq} &= N_0N_1 \Re\;[\textbf{DFT}(\widehat{T^{\bar{p}\bar{q}}} (\widehat{T^{pq}})^*)]
\end{split}
\end{equation}

Likewise, Equation~(\ref{eqn:sfft_eq23}) becomes
\begin{equation}
\label{eqn:B13}
\begin{split}
D_{\bar{i}\bar{j}\bar{\alpha}\bar{\beta}} &= \Theta_{\bar{i}\bar{j}}(\bar{\alpha}, \bar{\beta}) - \Theta_{\bar{i}\bar{j}} (0, 0) \;\;\;\textbf{\textit{if}} (\bar{\alpha}, \bar{\beta}) \neq \vec{0}
\\
D_{\bar{i}\bar{j}\bar{\alpha}\bar{\beta}} &= \Theta_{\bar{i}\bar{j}} (0, 0) \;\;\;\;\;\;\;\;\;\;\;\;\;\;\;\;\;\;\;\;\;\textbf{\textit{if}} (\bar{\alpha}, \bar{\beta}) = \vec{0},
\end{split}
\end{equation}
where
\begin{equation}
\label{eqn:B14}
\begin{split}
\Theta_{\bar{i}\bar{j}} (\rho, \epsilon) &= \mathring{\Theta}_{\bar{i}\bar{j}} (\rho\bmod{N_0}, \epsilon\bmod{N_1})
\\
\mathring{\Theta}_{\bar{i}\bar{j}} &= \Re\;[\textbf{DFT}(\widehat{S}^* (\widehat{I^{\bar{i}\bar{j}}})^*].
\end{split}
\end{equation}

Equation~(\ref{eqn:sfft_eq24}) at the right-hand-side becomes

\begin{equation}
\label{eqn:B15}
\begin{split}
E_{\bar{p}\bar{q}} &= \Delta_{\bar{p}\bar{q}} (0, 0),
\end{split}
\end{equation}
where
\begin{equation}
\label{eqn:B16}
\begin{split}
\Delta_{\bar{p}\bar{q}} (\rho, \epsilon) &= \mathring{\Delta}_{\bar{p}\bar{q}} (\rho\bmod{N_0}, \epsilon\bmod{N_1})
\\
\mathring{\Delta}_{\bar{p}\bar{q}} &= N_0N_1 \Re\;[\widehat{S}^* \widehat{T^{\bar{p}\bar{q}}}]
\end{split}
\end{equation}

Now the equations (\ref{eqn:sfft_eq19})-(\ref{eqn:sfft_eq24}) have been converted to equations (\ref{eqn:B5}), (\ref{eqn:B7}), (\ref{eqn:B9}), (\ref{eqn:B11}), (\ref{eqn:B13}), and (\ref{eqn:B15}). These equations used to establish the linear system are mainly comprised of DFTs and element-wise matrix operations, which allows for highly efficient parallel computations on GPU architecture.

\section{Noise Decorrelation} \label{App::noise-decorr}

\begin{figure*}[ht!]
    \centering
    \includegraphics[trim=1cm 0cm 1cm 0cm,clip=true,width=14cm]{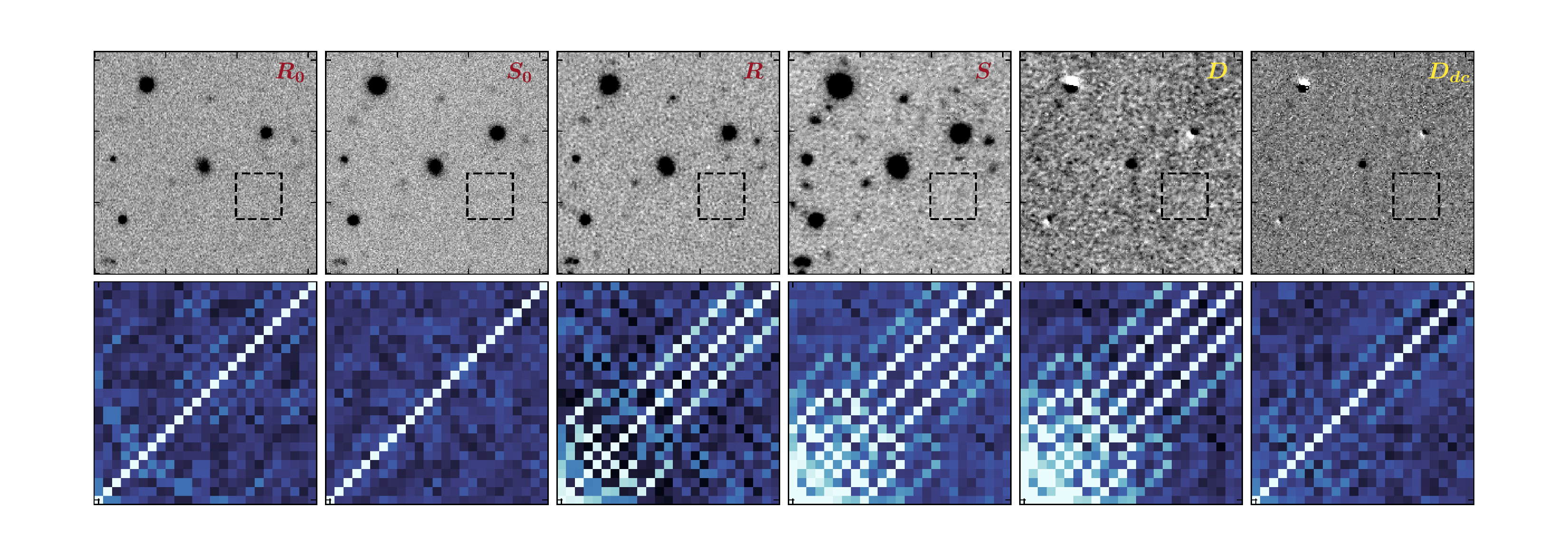}
    \caption{\label{fig:DeCorr} {Noise decorrelation test with DECam data. The upper panels show the zoomed-in region of SN candidate AT~2018fjd (RA=21:12:25.980 DEC=-66:00:56.50) discovered by DECamERON \citep{DECamERON_Mould17,AT2018fjd_Hu18}. $\text{R}_\text{0}$ ($\text{S}_\text{0}$) is from indivial observations $\textit{DECam-OBS18a}$ ($\textit{DECam-OBS04a}$). R (S) is the stacked reference (science) image. The subtraction between R and S results in the difference image D, and $\text{D}_\text{dc}$ is the decorrelated difference image. The black dashed boxes (41 $\times$ 41 pixels) are selected on the a proximate background region. The lower panel shows the corresponding covariance matrix measured from the boxes. To probe the local correlation of background pixels, we constructed a multivariate random variable $X$ for the flux of adjacent pixels with fixed positional relationship, where $X$ = ($X_{(0,0)}$, $X_{(1,0)}$, $X_{(-1,0)}$, $X_{(0,1)}$, $X_{(0,-1)}$, $X_{(1,1)}$, $X_{(1,-1)}$, $X_{(-1,1)}$, $X_{(-1,-1)}$, $X_{(2,0)}$, $X_{(-2,0)}$, $X_{(0,2)}$, $X_{(0,-2)}$, $X_{(3,0)}$, $X_{(-3,0)}$, $X_{(0,3)}$, $X_{(0,-3)}$, $X_{(4,0)}$, $X_{(-4,0)}$, $X_{(0,4)}$, $X_{(0,-4)}$, $X_{(5,0)}$, $X_{(-5,0)}$, $X_{(0,5)}$, $X_{(0,-5)}$). Note the subscripts represent the relative pixel locations with respect to the first element. Only the pixels within the boxes are allowed to be involved in the random vector. The covariance matrix is calculated by exhausting all possible realization of $X$.}}
\end{figure*}

\begin{deluxetable*}{llllllll}[ht!]
    \tablenum{A1}
    \tablewidth{0pt}
    \tablecolumns{4}
    \tablecaption{\label{tab:sfft_param} SFFT Parameters}
    
    \tablehead{
    \colhead{Parameter}           &
    \colhead{Value}               &
    \colhead{Note}               &
    \colhead{Comment}            &
    }
    
    \startdata
    $\text{-BeltHW}$    &     $\text{0.2}$    &     $\quad\text{Half-width of the }\textit{point-source-belt}$    &     $\text{only for }\textit{sparse-flavor}$
    \\
    $\text{-MAGD\_THRESH}$    &     $\text{0.12}$    &     $\quad\text{Outlier threshold to reject variable sources (mag)}$    &     $\text{only for }\textit{sparse-flavor}$
    \\
    $\text{-KerPolyOrder}$    &     $\text{2}$    &     $\quad\text{Spatial order of kernel variation}$    &     $\text{-}$
    \\
    $\text{-BGPolyOrder}$    &     $\text{2}$    &     $\quad\text{Spatial order of background variation}$    &     $\text{-}$
    \\
    $\text{-KerHWRatio}$    &     $\text{2}$    &     $\quad\text{The ratio between seeing and convolution kernel half-width}$    &     $\text{-}$
    \\
    $\text{-ConstPhotRatio}$    &     $\text{True}$    &     $\quad\text{Constant photometric ratio between reference and science}$    &     $\text{-}$
    \enddata
\end{deluxetable*}

Like in other subtraction methods emanated from \citet{AL98}, SFFT will introduce correlated errors on the difference image through the convolution process. 
For the cross-convolution approach, \citet{ZOGY} has proved that one can whiten the difference image by one convolution with a decorrelation kernel.
Later, \citet{LSST_Reiss16} presented a noise decorrelation strategy for image subtraction in the framework of \citet{AL98}. Here we extend to a more general case. That is, the subtraction can be performed with images constructed from the average or median co-addition of individual observations, and the original images are preprocessed by convolutions to match their PSFs before the co-additions.

The following derivation does not consider the spatial variation of convolution kernels across the field-of-view. However, it is appropriate to perform the operation by splitting the images into smaller segments to minimize the effect of PSF variation on correlated background noise. Likewise, the spatial variation of the background map for the input images is not considered. 
Note that a constant map has nothing to do with the noise correlation. Therefore, it is appropriate to assume that all the input images have zero mean background in our derivation.
Note that noise correlation caused by resampling for image alignment and drizzling is not considered here, as such operations cannot be simplified as convolution, and it is not easy to formulate how these procedures affect the statistical properties of the noise. 
Finally, we assume that the noise of the original images before convolution to be uncorrelated Poisson noise.

Given $M$ science images $S_j \sim \mathcal{N}(0, \sigma_{S_j}^{2})$ with $j$ being integers from 0, $M-1$, and reference images $R_i \sim \mathcal{N}(0, \sigma_{R_i}^{2})$ with $i$ being integers from 0, $N-1$, respectively. We consider the generic case of subsequent average or median co-additions produce the stacked science and reference images. The image subtraction employed to obtain the final difference image $D$ is given by

\begin{equation}
    \label{eqn:21}
    D = \sum_{j=0}^{M-1} S_{j} \circledast G_j - [\sum_{i=0}^{N-1} R_i \circledast H_i] \circledast K
\end{equation}

For simplicity, the convolution kernels $G_j$ and $H_i$ for PSF-homogenisation are divided by number of science ($S_j$) and reference ($R_i$) images $M$ and $N$, respectively. Assuming that a decorrelation kernel $Q$ exists which can whiten the noise of $D$, we can calculate $Q$ using the equation $\mathfrak{D} = D \circledast Q$ where $\mathfrak{D} \sim \mathcal{N}(0,\,\sigma_{\mathfrak{D}}^{2})$.

The covariance matrix of $\mathfrak{D}$ in Fourier space can be written as

\begin{equation}
\label{eqn:22}
\begin{split}
{\rm Cov}(\widehat{\mathfrak{D}}(p), \widehat{\mathfrak{D}}(q)) =  L_x & L_y \delta(p,q) \widehat{Q}(p) \widehat{Q}(q)^*
\\
\{\sum_j{\sigma_{S_j}^2 \widehat{G_j}(p) \widehat{G_j}(q)^*} &+ \sum_i{\sigma_{R_i}^2 \widehat{K}(p) \widehat{K}(q)^* \widehat{H_i}(p) \widehat{H_i}(q)^*}\},
\end{split}
\end{equation}
where $L_x$ and $L_y$ are the image dimensions in $x$ and $y$ directions, respectively, and
$p$ and $q$ are pixel indices of the image frame.

The Foruier Transform of the decorrelation kernel $\widehat{Q}$ can be derived by setting $q = p$,
\begin{equation}
\label{eqn:23}
\begin{split}
\abs{\widehat{Q}(p)}^2 = \sigma_{\mathfrak{D}}^2 / \{ \sum_j{\sigma_{S_j}^2 \abs{\widehat{G_j}(p)}^2} + \sum_i{\sigma_{R_i}^2 \abs{\widehat{K}(p)}^2 \abs{\widehat{H_i}(p)}^2} \}.
\end{split}
\end{equation}

The final solution to $\widehat{Q}$ should be conjugate symmetric to ensure its inverse Fourier Transformation to be real. Here we consider the most straightforward solution with $\widehat{Q}$ being a real function only, that is

\begin{equation}
\label{eqn:24}
\begin{split}
\widehat{Q_0} = \sqrt{\sigma_{\mathfrak{D}}^2 / \{ \sum_j{\sigma_{S_j}^2 \abs{\widehat{G_j}}^2} + \sum_i{\sigma_{R_i}^2 \abs{\widehat{K}}^2 \abs{\widehat{H_i}}^2} \}}.
\end{split}
\end{equation}

Note $\sigma_{\mathfrak{D}}$ is the desired noise of decorrelated difference image. In principle, it should be set to be as close as possible to the true background in regions free of astronomical objects. In practice, this normalization can also be set by requiring the decorrelation procedure to preserve the flux zero-point of the images. In the latter case, SFFT simply normalizes the decorrelation kernel $Q$ to have a unit kernel sum.

There is a major difference between SFFT and ZOGY \citep{ZOGY} for noise decorrelation, although they share the same statistical principles. In ZOGY, convolution kernels for co-addition and subtraction are tied to the PSFs of the images. SFFT does not need to model the PSFs for the images $S_j$ and $R_i$ for the noise decorrelation process. Only the match kernels $G_j$, $H_i$, and $K$ are needed for SFFT.
This can be even more convenient for images with simple co-additions of original observations without matching the PSFs, such as observations taken under stable seeing conditions, only the match kernel $K$ is needed for SFFT, and no PSF construction is necessary.  

We select ten DECam images from Table~\ref{tab:test_data_info} for testing the performance of noise decorrelation. 
The generic case of image subtraction follows Equation~(\ref{eqn:21}), with \{$R_0$, $R_1$, $R_2$, $R_3$, $R_4$\} = \{$\textit{DECam-OBS18a}$, $\textit{DECam-OBS18b}$, $\textit{DECam-OBS18c}$, $\textit{DECam-OBS18d}$, $\textit{DECam-OBS18e}$\} being the individual reference images, and \{$S_0$, $S_1$, $S_2$, $S_3$, $S_4$\} = \{$\textit{DECam-OBS04a}$, $\textit{DECam-OBS04b}$, $\textit{DECam-OBS04c}$, $\textit{DECam-OBS04d}$, $\textit{DECam-OBS04e}$\} being the individual science images. All the images have been astronomically aligned to $S_0$ by {\tt\string SWarp}.
We create the stacked reference image by median co-addition of the five individual reference images with PSF homogenization matching to $R_0$. Similarly, the stacked science image is the median coadd of the five individual science images after matching their PSFs to $S_0$. The final subtraction is performed between the stacked reference and stacked science image. 
Noise decorrelation for the difference image is conducted by convolving the decorrelation kernel calculated by Equation~(\ref{eqn:24}).
As the spatial variation has been ignored, here, the example only focuses on a small area around the supernova candidate AT 2018fjd discovered in DECamERON survey \citep{AT2018fjd_Hu18}.

We trace the background pixel correlation along with the coadd and subtraction processes, shown in Figure~\ref{fig:DeCorr}. Note that image alignment can also introduce pixel correlation via a localized resampling function (LANCZOS3 function of {\tt\string SWarp} in this case). One can notice that the pixel correlation in resampled image $R_0$ is already slightly higher than un-resampled image $S_0$.
The correlation is increased drastically in co-added images as multiple convolutions for PSF homogenization come into play in this process.
It becomes even stronger after the final subtraction. Nevertheless, our decorrelation scenario can effectively reduce the noise correlation on the final difference image to a quite low level comparable with that of a resampled individual image $R_0$.

\begin{deluxetable*}{llllllll}[ht!]
    \tablenum{A2}
    \tablewidth{0pt}
    \tablecolumns{3}
    \tablecaption{\label{tab:hp_param} {\tt\string HOTPANTS} Parameters}
    
    \tablehead{
    \colhead{Parameter}           &
    \colhead{Value}               &
    \colhead{Note}               &
    }
    
    \startdata
    $\text{-r}$    &     $\quad\text{2}\times\text{FWHM}_\text{L}$    &     $\text{Convolution kernel half-width (pixel)}$
    \\
    $\text{-rss}$    &     $\quad\text{6}\times\text{FWHM}_\text{L}$    &     $\text{Substamp half-width (pixel)}$
    \\
    $\text{-nsx}$    &     $\quad\text{NX / 200}$    &     $\text{Image segmentation of stamps (x axis)}$
    \\
    $\text{-nsy}$    &     $\quad\text{NY / 200}$    &     $\text{Image segmentation of stamps (y axis)}$
    \\
    $\text{-ko}$    &     $\quad\text{2}$    &     $\text{Spatial order of kernel variation}$
    \\
    $\text{-bgo}$    &     $\quad\text{2}$    &     $\text{Spatial order of background variation}$
    \\
    $\text{-tu}$    &     $\quad\text{saturation recorded in reference header}$    &     $\text{Upper valid flux of the reference image}$
    \\
    $\text{-tl}$    &     $\quad\text{sky(reference)}-\text{10}\times\text{skysigma(reference)}$    &     $\text{Lower valid flux of the reference image}$
    \\
    $\text{-iu}$    &     $\quad\text{saturation recorded in science header}$    &     $\text{Upper valid flux of the science image}$
    \\
    $\text{-il}$    &     $\quad\text{sky(science)}-\text{10}\times\text{skysigma(science)}$    &     $\text{Lower valid flux of the science image}$
    \enddata
    \tablecomments{$\text{FWHM}_\text{L}$ is the worse seeing measured from the input image-pair. Note the ratio of 2 in parameter $-r$ is consistent with the default configuration of SFFT (see $\textbf{\text{-KerHWRatio}}$). The parameters $\text{-ko}$ and $\text{-bgo}$ are equivalent to SFFT paramters $\textbf{\text{-KerPolyOrder}}$ and $\textbf{\text{-BGPolyOrder}}$, respectively. The symbols NX and NY indicate the image size along x axis and y axis, respectively. sky and skysigma describe the sky background measured by MMM algorithm. The {\tt\string HOTPANTS} configurations here largely refer to those in iPTF image subtraction pipeline \citep{iPTF_Cao16}.}
\end{deluxetable*}

\begin{deluxetable*}{llllllll}[ht!]
    \tablenum{A3}
    \tablewidth{0pt}
    \tablecolumns{3}
    \tablecaption{\label{tab:zogy_param} {\tt\string ZOGY} Parameters}
    
    \tablehead{
    \colhead{Parameter}           &
    \colhead{Value}               &
    \colhead{Note}               &
    }
    
    \startdata
    $\text{-subimage\_size}$    &     $\quad\text{256}$    &     $\text{Size of subimages (pixel)}$
    \\
    $\text{-subimage\_border}$    &     $\quad\text{32}$    &     $\text{Border around subimage to avoid edge effects (pixel)}$
    \\
    $\text{-fratio\_local}$    &     $\quad\text{F}$    &     $\text{Flux ratio from subimage (T) or full frame (F)}$
    \\
    $\text{-psf\_poldeg}$    &     $\quad\text{2}$    &     $\text{Polynomial degree for PSF spatial variation}$
    \\
    $\text{-size\_vignet}$    &     $\quad\text{49}$    &     $\text{Size of the {\tt  \string SExtractor} VIGNETS for PSF construction}$
    \\
    $\text{-nthreads}$    &     $\quad\text{32}$    &     $\text{Number for CPU multithreading}$
    \enddata
\end{deluxetable*}

\begin{deluxetable*}{llllllll}[ht!]
    \tablenum{A4}
    \tablewidth{0pt}
    \tablecolumns{3}
    \tablecaption{\label{tab:ptdia_param} {\tt\string PyTorchDIA} Parameters}
    
    \tablehead{
    \colhead{Parameter}           &
    \colhead{Value}               &
    \colhead{Note}               &
    }
    
    \startdata
    $\text{-loss\_fn}$    &     $\quad\text{robust}$    &     $\text{loss function (robust or Gaussian)}$
    \\
    $\text{-flat}$    &     $\quad\text{1}$    &     $\text{Initialization value of the convolution kernel}$
    \\
    $\text{-ks}$    &     $\quad\text{4}\times\text{FWHM}_\text{L}+1$    &     $\text{Kernel size}$
    \\
    $\text{-poly\_degree}$    &     $\quad\text{2}$    &     $\text{Polynomial degree for background fit}$
    \enddata
    \tablecomments{$\text{FWHM}_\text{L}$ is the worse seeing measured from the input image-pair.}
\end{deluxetable*}

\begin{deluxetable*}{llllllll}[ht!]
    \tablenum{B}
    \tablewidth{0pt}
    \tablecolumns{6}
    \tablecaption{\label{tab:test_data_info} Specification of the test data}
    
    \tablehead{
    \colhead{Instrument}               &
    \colhead{File Alias}               &
    \colhead{File Name}                &
    \colhead{Exposure Time}            &
    \colhead{FWHM}                     &
    \colhead{5$\sigma$ Lim. Mag}       &
    \\
    \colhead{}                         &
    \colhead{}                         &
    \colhead{}                         &
    \colhead{(s)}                      &
    \colhead{(arcsec)}                 &
    \colhead{(mag)}
    }
    
    \startdata
    $\,\,\,\text{ZTF}$    &    $\,\,\,\textit{ZTF-REF}$    &   $\,\,\,\text{ztf$\_$001735$\_$zg$\_$c01$\_$q2$\_$refimg}$     &   $\quad\quad\text{1200}$   &   $\,\,\,\text{2.02}$    &   $\quad\,\,\,\text{22.89}$
    \\
    $\,\,\,\text{ZTF}$    &    $\,\,\,\textit{ZTF-SCI}$    &   $\,\,\,\text{ztf$\_$20180705481609$\_$001735$\_$zg$\_$c01$\_$o$\_$q2$\_$sciimg}$     &   $\quad\quad\text{30}$   &   $\,\,\,\text{1.80}$    &   $\quad\,\,\,\text{20.68}$
    \\
    $\,\,\,\text{AST3-II}$    &    $\,\,\,\textit{AST-REF}$    &   $\,\,\,\text{a0331.94}$     &   $\quad\quad\text{60}$   &   $\,\,\,\text{2.41}$    &   $\quad\,\,\,\text{-}$
    \\
    $\,\,\,\text{AST3-II}$    &    $\,\,\,\textit{AST-SCI}$    &   $\,\,\,\text{a0405.162}$     &   $\quad\quad\text{60}$   &   $\,\,\,\text{2.96}$    &   $\quad\,\,\,\text{-}$
    \\
    $\,\,\,\text{TESS}$    &    $\,\,\,\textit{TESS-REF}$    &   $\,\,\,\text{tess2018258205941-s0002-1-1-0121-s$\_$ffic.001}$     &   $\quad\quad\text{1440}$   &   $\,\,\,\text{36.95}$    &   $\quad\,\,\,\text{-}$
    \\
    $\,\,\,\text{TESS}$    &    $\,\,\,\textit{TESS-SCI}$    &   $\,\,\,\text{tess2018258225941-s0002-1-1-0121-s$\_$ffic.001}$     &   $\quad\quad\text{1440}$   &   $\,\,\,\text{36.89}$    &   $\quad\,\,\,\text{-}$
    \\
    $\,\,\,\text{DECam}$    &    $\,\,\,\textit{DECam-SREF}$    &   $\,\,\,\text{c4d$\_$180806$\_$052738$\_$ooi$\_$i$\_$v1}$     &   $\quad\quad\text{330}$   &   $\,\,\,\text{0.85}$    &   $\quad\,\,\,\text{24.35}$
    \\
    $\,\,\,\text{DECam}$    &    $\,\,\,\textit{DECam-PTAR}$    &   $\,\,\,\text{c4d$\_$150821$\_$001224$\_$ooi$\_$i$\_$v1}$     &   $\quad\quad\text{330}$   &   $\,\,\,\text{1.40}$    &   $\quad\,\,\,\text{22.73}$
    \\
    $\,\,\,\text{DECam}$    &    $\,\,\,\textit{DECam-OBS18a}$    &   $\,\,\,\text{c4d$\_$160818$\_$043331$\_$ooi$\_$i$\_$v1}$     &   $\quad\quad\text{330}$   &   $\,\,\,\text{1.04}$    &   $\quad\,\,\,\text{23.28}$
    \\
    $\,\,\,\text{DECam}$    &    $\,\,\,\textit{DECam-OBS18b}$    &   $\,\,\,\text{c4d$\_$160818$\_$020339$\_$ooi$\_$i$\_$v1}$     &   $\quad\quad\text{330}$   &   $\,\,\,\text{1.07}$    &   $\quad\,\,\,\text{23.35}$
    \\
    $\,\,\,\text{DECam}$    &    $\,\,\,\textit{DECam-OBS18c}$    &   $\,\,\,\text{c4d$\_$160818$\_$030325$\_$ooi$\_$i$\_$v1}$     &   $\quad\quad\text{330}$   &   $\,\,\,\text{0.96}$    &   $\quad\,\,\,\text{23.45}$
    \\
    $\,\,\,\text{DECam}$    &    $\,\,\,\textit{DECam-OBS18d}$    &   $\,\,\,\text{c4d$\_$160818$\_$053313$\_$ooi$\_$i$\_$v1}$     &   $\quad\quad\text{330}$   &   $\,\,\,\text{1.12}$    &   $\quad\,\,\,\text{23.09}$
    \\
    $\,\,\,\text{DECam}$    &    $\,\,\,\textit{DECam-OBS18e}$    &   $\,\,\,\text{c4d$\_$160818$\_$065148$\_$ooi$\_$i$\_$v1}$     &   $\quad\quad\text{330}$   &   $\,\,\,\text{1.01}$    &   $\quad\,\,\,\text{23.32}$
    \\
    $\,\,\,\text{DECam}$    &    $\,\,\,\textit{DECam-OBS02a}$    &   $\,\,\,\text{c4d$\_$180802$\_$052807$\_$ooi$\_$i$\_$v1}$     &   $\quad\quad\text{330}$   &   $\,\,\,\text{1.38}$    &   $\quad\,\,\,\text{23.40}$
    \\
    $\,\,\,\text{DECam}$    &    $\,\,\,\textit{DECam-OBS02b}$    &   $\,\,\,\text{c4d$\_$180802$\_$055800$\_$ooi$\_$i$\_$v1}$     &   $\quad\quad\text{330}$   &   $\,\,\,\text{1.31}$    &   $\quad\,\,\,\text{23.55}$
    \\
    $\,\,\,\text{DECam}$    &    $\,\,\,\textit{DECam-OBS02c}$    &   $\,\,\,\text{c4d$\_$180802$\_$062754$\_$ooi$\_$i$\_$v1}$     &   $\quad\quad\text{330}$   &   $\,\,\,\text{1.22}$    &   $\quad\,\,\,\text{23.60}$
    \\
    $\,\,\,\text{DECam}$    &    $\,\,\,\textit{DECam-OBS02d}$    &   $\,\,\,\text{c4d$\_$180802$\_$065746$\_$ooi$\_$i$\_$v1}$     &   $\quad\quad\text{330}$   &   $\,\,\,\text{1.41}$    &   $\quad\,\,\,\text{23.33}$
    \\
    $\,\,\,\text{DECam}$    &    $\,\,\,\textit{DECam-OBS02e}$    &   $\,\,\,\text{c4d$\_$180802$\_$072740$\_$ooi$\_$i$\_$v1}$     &   $\quad\quad\text{330}$   &   $\,\,\,\text{1.21}$    &   $\quad\,\,\,\text{23.58}$
    \\
    $\,\,\,\text{DECam}$    &    $\,\,\,\textit{DECam-OBS02f}$    &   $\,\,\,\text{c4d$\_$180802$\_$075732$\_$ooi$\_$i$\_$v1}$     &   $\quad\quad\text{330}$   &   $\,\,\,\text{1.29}$    &   $\quad\,\,\,\text{23.62}$
    \\
    $\,\,\,\text{DECam}$    &    $\,\,\,\textit{DECam-OBS02g}$    &   $\,\,\,\text{c4d$\_$180802$\_$085719$\_$ooi$\_$i$\_$v1}$     &   $\quad\quad\text{330}$   &   $\,\,\,\text{1.24}$    &   $\quad\,\,\,\text{23.55}$
    \\
    $\,\,\,\text{DECam}$    &    $\,\,\,\textit{DECam-OBS02h}$    &   $\,\,\,\text{c4d$\_$180802$\_$092739$\_$ooi$\_$i$\_$v1}$     &   $\quad\quad\text{330}$   &   $\,\,\,\text{1.35}$    &   $\quad\,\,\,\text{23.42}$
    \\
    $\,\,\,\text{DECam}$    &    $\,\,\,\textit{DECam-OBS04a}$    &   $\,\,\,\text{c4d$\_$180804$\_$092331$\_$ooi$\_$i$\_$v1}$     &   $\quad\quad\text{330}$   &   $\,\,\,\text{1.41}$    &   $\quad\,\,\,\text{23.50}$
    \\
    $\,\,\,\text{DECam}$    &    $\,\,\,\textit{DECam-OBS04b}$    &   $\,\,\,\text{c4d$\_$180804$\_$052409$\_$ooi$\_$i$\_$v1}$     &   $\quad\quad\text{330}$   &   $\,\,\,\text{1.37}$    &   $\quad\,\,\,\text{23.62}$
    \\
    $\,\,\,\text{DECam}$    &    $\,\,\,\textit{DECam-OBS04c}$    &   $\,\,\,\text{c4d$\_$180804$\_$055402$\_$ooi$\_$i$\_$v1}$     &   $\quad\quad\text{330}$   &   $\,\,\,\text{1.57}$    &   $\quad\,\,\,\text{23.43}$
    \\
    $\,\,\,\text{DECam}$    &    $\,\,\,\textit{DECam-OBS04d}$    &   $\,\,\,\text{c4d$\_$180804$\_$075344$\_$ooi$\_$i$\_$v1}$     &   $\quad\quad\text{330}$   &   $\,\,\,\text{1.36}$    &   $\quad\,\,\,\text{23.61}$
    \\
    $\,\,\,\text{DECam}$    &    $\,\,\,\textit{DECam-OBS04e}$    &   $\,\,\,\text{c4d$\_$180804$\_$082338$\_$ooi$\_$i$\_$v1}$     &   $\quad\quad\text{330}$   &   $\,\,\,\text{1.10}$    &   $\quad\,\,\,\text{23.89}$
    \\
    $\,\,\,\text{DECam}$    &    $\,\,\,\textit{DECam-OBS04f}$    &   $\,\,\,\text{c4d$\_$180804$\_$045412$\_$ooi$\_$i$\_$v1}$     &   $\quad\quad\text{330}$   &   $\,\,\,\text{1.46}$    &   $\quad\,\,\,\text{23.45}$
    \\
    $\,\,\,\text{DECam}$    &    $\,\,\,\textit{DECam-OBS04g}$    &   $\,\,\,\text{c4d$\_$180804$\_$062356$\_$ooi$\_$i$\_$v1}$     &   $\quad\quad\text{330}$   &   $\,\,\,\text{1.68}$    &   $\quad\,\,\,\text{23.35}$
    \\
    $\,\,\,\text{DECam}$    &    $\,\,\,\textit{DECam-OBS04h}$    &   $\,\,\,\text{c4d$\_$180804$\_$065350$\_$ooi$\_$i$\_$v1}$     &   $\quad\quad\text{330}$   &   $\,\,\,\text{1.45}$    &   $\quad\,\,\,\text{23.58}$
    \\
    $\,\,\,\text{DECam}$    &    $\,\,\,\textit{DECam-OBS04i}$    &   $\,\,\,\text{c4d$\_$180804$\_$072342$\_$ooi$\_$i$\_$v1}$     &   $\quad\quad\text{330}$   &   $\,\,\,\text{1.66}$    &   $\quad\,\,\,\text{23.22}$
    \\
    $\,\,\,\text{DECam}$    &    $\,\,\,\textit{DECam-OBS04j}$    &   $\,\,\,\text{c4d$\_$180804$\_$085332$\_$ooi$\_$i$\_$v1}$     &   $\quad\quad\text{330}$   &   $\,\,\,\text{1.40}$    &   $\quad\,\,\,\text{23.51}$
    \\
    $\,\,\,\text{TMTS}$    &    $\,\,\,\textit{TMTS-REF}$    &   $\,\,\,\text{f20191103$\_$1$\_$NT0023$\_$L$\_$1965$\_$1970}$     &   $\quad\quad\text{60}$   &   $\,\,\,\text{6.45}$    &   $\quad\,\,\,\text{-}$
    \\
    $\,\,\,\text{TMTS}$    &    $\,\,\,\textit{TMTS-SCI}$    &   $\,\,\,\text{f20191028$\_$1$\_$NT0023$\_$L$\_$755$\_$760}$     &   $\quad\quad\text{60}$   &   $\,\,\,\text{7.46 }$    &   $\quad\,\,\,\text{-}$
    \enddata
    \tablecomments{$\textit{ZTF-REF}$, directly retreived from ZTF Data Release 3, is coadded from 40 individual observations as a deep reference. $\textit{TESS-REF}$ and $\textit{TESS-SCI}$ are Full-Frame images (FFIs) recorded at 30 minutes cadence, with effective integration time of 1440s. $\textit{TMTS-REF}$ ($\textit{TMTS-SCI}$) is median stacked from 6 consecutive observations, each of which has exposure time of 10 seconds. Other images in the table are single-exposure observations. Each DECam exposure is comprised of 62 CCD images, the FWHM and limiting magnitude for each exposure in the table shows the median measurement over all CCD tiles.}
\end{deluxetable*}


\bibliography{sfft}{}
\bibliographystyle{aasjournal}

\end{document}